\documentclass[12pt]{article}
\usepackage{latexsym,epsfig,graphicx}

\title{Kerr-Schild Method and Geodesic Structure in Codimension-2 Brane Black Holes}
\author{Bertha Cuadros-Melgar\footnote{e-mail: berthaki@gmail.com}\\
\normalsize{Departamento de Ciencias F\'{i}sicas, Universidad Andr\'es Bello,}\\
\normalsize{Avenida Rep\'ublica 252, Santiago, Chile}\\ \\
Susana Aguilar\footnote{e-mail: saguilar@ing.uchile.cl}~ and Nelson Zamorano\footnote{e-mail: nzamora@dfi.uchile.cl}\\
\normalsize{Departamento de F\'{i}sica, Facultad de Ciencias F\'{i}sicas y Matem\'aticas,} \\
\normalsize{Universidad de Chile, Avenida Blanco Encalada 2003, Santiago, Chile}}
\date{}

\begin{document}\maketitle

\begin{abstract}
We consider the black hole solutions of five-dimensional gravity with a Gauss-Bonnet term in the bulk and an induced gravity term on a 2-brane of codimension-2. Applying the Kerr-Schild method we derive additional solutions which include charge and angular momentum. To improve our understanding of this geometry we study the geodesic structure of such spacetimes.
\end{abstract}

\begin{center}
\end{center}

\newpage
\section{Introduction}

One of the unsolved questions of braneworld cosmology is the existence of localized black holes on the brane. This puzzle has been investigated almost since the appearance of these alternative-to-classical-gravity models. In codimension-1 scenarios the natural first proposal was to consider the Schwarzschild metric and study its black string extension into the bulk~\cite{Chamblin:1999by}. Unfortunately, as intuited by the authors, this string is unstable to classical linear perturbations known as Gregory-Laflamme (GL) instability~\cite{BSINS}. Since then there has been an intensive research to find a full metric by using numerical techniques~\cite{BHNUM} or by solving the trace of the projected Einstein equations on the brane~\cite{SMS} with special Ansatz~\cite{Cas-Bro} or making certain assumptions on the projected Weyl term coming from the bulk~\cite{BBH}. A lower dimensional version with a Ba\~nados-Teitelboim-Zanelli (BTZ)~\cite{Banados:1992wn} black string was also considered in~\cite{EHM}. This solution was obtained from the so-called C-metric. The thermodynamical analysis showed that the string remains stable when its transverse size is comparable to the 4-dimensional AdS radius and can be unbalanced by a GL instability above that scale, breaking up to a BTZ brane black hole.

In codimension-2 the first attempt was proposed in~\cite{Kaloper:2006ek} as a generalization of the
4-dimensional Aryal, Ford, Vilenkin black hole~\cite{Aryal:1986sz} pierced by a cosmic string. The
rotating version was also presented in~\cite{Kiley:2007wb}, and a complete study
of the gray-body factors was considered in~\cite{dejan}. Another proposal came by considering a
5-dimensional bulk with a Gauss-Bonnet term and induced gravity on the brane~\cite{CuadrosMelgar:2007jx}.
 The solutions are basically of three types. The first one is the familiar BTZ black hole which can be extended into the bulk with a regular horizon. The second one adds a short distance correction and describes a BTZ black hole conformally coupled to a scalar field. There is a third solution family that can accommodate any brane metric coefficient $n(r)$ provided that it can yield a physically acceptable brane energy-momentum tensor. The corresponding generalization to 6-dimensional bulks was worked out in~\cite{cptz2}, where it is shown that the only possible solution, namely a Schwarzschild-AdS brane black hole, needs matter in the
bulk. For all these solutions the Gauss-Bonnet term plays a fundamental r\^ole leading to a consistency
relation that dictates the kind of bulk or brane matter necessary to sustain a black hole on the brane.

The arbitrariness of $n(r)$ in one of the above mentioned solutions is an interesting feature that
deserves more discussion. This fact motivated us to look for more solutions that could exhibit physically
 acceptable brane energy-momentum tensors. Specifically, we wondered if charged and rotating
 metrics can be included in this general $n(r)$, in particular, taking into account that the rotating
 BTZ black hole has non-diagonal elements that were not included in the original derivation of the
 solutions found in~\cite{CuadrosMelgar:2007jx}. With this purpose we recur to the so-called
 Kerr-Schild method.

The Kerr-Schild coordinates appeared for the first time when obtaining the Kerr metric starting from
a flat empty space. Later on, A.~H.~Taub~\cite{taub} generalized this method in order to obtain new
solutions by adding to a background metric a term proportional to a null geodesic vector and a scalar
function $H$. The resulting metric is not a coordinate transformation but a new spacetime with a
different geometry. The selection of $H$, although arbitrary, must fulfill certain criteria. Not only
it has to satisfy the Einstein equations when inserted in the new metric but it must also yield a
 physically meaningful energy-momentum tensor. The advantage of this Generalized Kerr-Schild (GKS)
 method is that the new Einstein equations are linear in $H$ when written in their covariant and
 contravariant components. We should also mention that the resulting metric can have non-diagonal
 terms that can be cast off by means of a coordinate transformation when no angular momentum is
 involved. The GKS transformation appears as a useful tool in astrophysics where, for example, it can
 generate a singularity free metric that describes orbits close to a Kerr black hole event
 horizon~\cite{kom}. 
Moreover, this kind of metric can be applied in numerical relativity to outline the geometry at the horizon and even extend it to the interior of the black hole, given the absence of singularities of these coordinates in this region. The GKS transformation is also used to
 identify apparent horizons in boosted black holes due to its invariance under Lorentz
 boost~\cite{huq}. Other solutions were also worked out in~\cite{zam1}. This method has also been used to find exact vacuum solutions in the context of multidimensional gravity~\cite{Anabalon}.

In this paper we applied the GKS method to extra dimensions. The background solutions are the black
hole metrics described previously. Due to the Gauss-Bonnet contribution, the Einstein equations are
not linear anymore even when written in their covariant-contravariant form, in fact, they are at most quadratic in the function $H$. However, as we will see
in this work, the method proved to be successful since the equations are still solvable, and it was possible to obtain more solutions
including charge and angular momentum. It was also viable to pass from BTZ metrics (pure or charged)
to the so-called corrected BTZ one, which includes a brane scalar field.

Given all these solutions it is pertinent to explore the main properties of such spacetimes.
In General Relativity one possible way to understand the geometrical aspect of the gravitational field is to study
geodesics in the spacetime permeated by this field. Over the years, the motion of massive and
massless test particles in background geometries of various higher dimensional theories of
gravity has been investigated~\cite{highgeo}. In this work we analyse the timelike and null
geodesic motion in the brane black hole spacetimes mentioned above. With this aim we solve
the Euler-Lagrange equation for the variational problem associated with the corresponding
metrics. The set of orbits turned out to be very rich.

The paper is organized as follows. In section 2 we make a brief review of the brane black holes obtained in~\cite{CuadrosMelgar:2007jx}. Section 3 is devoted to the application of the GKS method to these models displaying the corresponding results and new solutions. In section 4 we present a complete study of timelike and null geodesic behaviour in the whole family of backgrounds. Finally, in section 5 we discuss our results and conclude.

\section{BTZ String-like Solutions on Codimension-2 Braneworlds}

We consider the following gravitational action in five dimensions with a Gauss-Bonnet (GB) term in the bulk and an induced three-dimensional curvature term on the brane~\cite{CuadrosMelgar:2007jx}
\begin{eqnarray}
\label{AcGBIG}
S_{\rm grav}=\left\{ \int d^5 x\sqrt{-g^{(5)}}\left[ R^{(5)}
+\alpha\left( R^{(5)2}-4 R^{(5)}_{MN}R^{(5)MN}+R^{(5)}_{MNKL}R^{(5)MNKL}\right)\right] \right.\nonumber\\
+ \left. r^{2}_{c} \int d^3x\sqrt{- g^{(3)}}\,R^{(3)}\right\}\frac{M^{3}_{5}}{2}+\int d^5 x \mathcal{L}_{bulk}+\int d^3 x \mathcal{L}_{brane}\,, \qquad \qquad \quad
\label{5daction}
\end{eqnarray}
where $\alpha\, (\geq0)$ is the GB coupling constant, $r_c ^2=M_{3}/M_{5}^3$ is the induced gravity ``cross-over" scale, which marks the transition from 3D to 5D gravity, and $M_{5}$, $M_{3}$ are the five and three-dimensional Planck masses, respectively.

The above induced term has been written in the particular coordinate system in which the metric is
\begin{equation}
ds_5^2=g_{\mu\nu}(x,\rho)dx^\mu
dx^\nu+d\rho^2+b^2(x,\rho)d\theta^2~.\label{5dmetric}
\end{equation}
Here $g_{\mu\nu}(x,0)$ is the brane metric, whereas $x^{\mu}$ denotes three  dimensions, $\mu=t,r,\phi$, and $\rho,\theta$ denote the radial and angular coordinates of the two extra dimensions ($\rho$ may or may not be compact, and $0\leq \theta < 2\pi$).
Capital $M$,~$N$ indices will take values in the five-dimensional space.

The Einstein equations resulting from the variation of the action~(\ref{5daction}) are
\begin{equation}
G^{(5)N}_M + r_c^2 G^{(3)\nu}_\mu g_M^\mu g^N_\nu {\delta(\rho) \over 2 \pi b}-\alpha H_{M}^{N} =\frac{1}{M^{3}_{5}} \left[T^{(B)N}_M+T^{(br)\nu}_\mu g_M^\mu g^N_\nu {\delta(\rho) \over 2 \pi b}\right]~,
\label{einsequat3}
\end{equation}
where
\begin{eqnarray}
H_M^N&=& \left[{1 \over 2}g_M^N (R^{(5)~2} -4R^{(5)~2}_{KL}+R^{(5)~2}_{ABKL})\right.-2R^{(5)}R^{(5)N}_{M}\nonumber\\
&&+4R^{(5)}_{MP}R^{NP}_{(5)}\phantom{{1 \over 2}}~\left. +4R^{(5)~~~N}_{KMP}R_{(5)}^{KP} -2R^{(5)}_{MKL P}R_{(5)}^{NKLP}\right]~.
\label{gaussbonnet}
\end{eqnarray}

To obtain the braneworld equations we expand the metric around the brane as
\begin{equation}
b(x,\rho)=\beta(x)\rho+O(\rho^{2})~.
\end{equation}
At the boundary of the internal two-dimensional space where the 2-brane is situated the function $b$ behaves as $b^{\prime}(x,0)=\beta(x)$, where a prime
denotes derivative with respect to $\rho$. In addition,
we demand that the space in the vicinity of the conical singularity is regular, {\it i.e.}, $\partial_\mu \beta=0$ and $\partial_{\rho}g_{\mu\nu}(x,0)=0$~\cite{Bostock:2003cv}.
The extrinsic curvature in the particular gauge $g_{\rho \rho}=1$ that we are considering  is given by $K_{\mu\nu}=g'_{\mu\nu}$.
Using the fact that the second derivatives of the metric contain $\delta$-function singularities at the position of the brane, the nature of the singularity gives the following relations \cite{Bostock:2003cv}
\begin{eqnarray}
{b'' \over b}&=&-(1-b'){\delta(\rho) \over b}+ {\rm non-singular~terms}~,\\
{K'_{\mu\nu} \over b}&=&K_{\mu\nu}{\delta(\rho) \over b}+ {\rm non-singular~terms}~.
\end{eqnarray}
From the above singularity expressions and using the Gauss-Codacci equations, we can  match the singular parts of the Einstein equations (\ref{einsequat3}) and get the ``boundary" Einstein equations,
\begin{equation}
G^{(3)}_{\mu\nu}={1 \over M_{3}^2} T^{(br)}_{\mu\nu}+2\pi (1-\beta) {M_{5}^3 \over M_{3}^2}g_{\mu\nu}
\label{einsteincomb3}~.
\end{equation}

We look for black string solutions of the Einstein equations~(\ref{einsequat3}) using the five-dimensional metric~(\ref{5dmetric}) in the form
\begin{equation}
ds_5^2=f^{2}(\rho)\left[-n(r)^{2}dt^{2}+n(r)^{-2}dr^{2}+r^{2} d\phi^{2}\right]+d\rho^2+b^2(\rho)d\theta^2~,
\label{5smetricc}
\end{equation}
where we have supposed the existence of a localized (2+1) black hole on the brane, whose metric is given by
\begin{equation}
ds_{3}^{2}=-n(r)^{2}dt^{2}+n(r)^{-2}dr^{2}+r^{2}d\phi^{2}~.
\label{3dmetric}
\end{equation}

In the bulk we consider only a cosmological constant $\Lambda_{5}$. Then, from the bulk Einstein equations
\begin{equation}
G^{(5)}_{MN}-\alpha
H_{MN}=-\frac{\Lambda_{5}}{M^{3}_{5}}g_{MN}~.
\end{equation}
By combining the $(rr,\phi \phi)$ equations we get
\begin{equation}
\left(\dot{n}^{2}+n \ddot{n}-\frac{n \dot{n}}{r}\right)\left(1-4\alpha \frac{b''}{b}\right)=0~,\label{173}
\end{equation}
while a combination of the $(\rho \rho, \theta \theta)$ equations gives
\begin{equation}
\left(f''-\frac{f'b'}{b}\right)\left[3-4\frac{\alpha}{f^{2}}\left(\dot{n}^{2}+n
\ddot{n}+2\frac{n \dot{n}}{r}+3f'^{2} \right)\right]=0\label{183}~,
\end{equation}
where a  dot implies derivatives with respect to $r$. The solutions of the equations (\ref{173}) and (\ref{183}) are summarized in Table \ref{table1}~\cite{CuadrosMelgar:2007jx}.

\begin{table}[here]
\begin{center}
\begin{tabular}{cccccc}
\hline
  $n(r)$ & $f(\rho)$ & $b(\rho)$ & $-\Lambda_5$ & Constraints \\
\hline
  BTZ & $\cosh\left(\frac{\rho}{2\,\sqrt{\alpha}}\right)$ & $\forall b(\rho)$ &
$\frac{3}{4\alpha}$ &
$l^2=4\,\alpha$ \\
  BTZ & $\cosh\left(\frac{\rho}{2\,\sqrt{\alpha}}\right)$ & $2\,\beta\,\sqrt{\alpha}\,\sinh\left(\frac{\rho}{2\,\sqrt{\alpha}}\right)$ &
  $\frac{3}{4\alpha}$ & - \\
  BTZ & $\cosh\left(\frac{\rho}{2\,\sqrt{\alpha}}\right)$ & $2\,\beta\,\sqrt{\alpha}\,\sinh\left(\frac{\rho}{2\,\sqrt{\alpha}}\right)$ &
  $\frac{3}{4\alpha}$ & $l^2=4\,\alpha$ \\
  BTZ & $\pm 1$ & $\gamma\,\sinh\left(\rho/\gamma\right)$ & $\frac{3}{l^2}$ & $\gamma=\sqrt{\frac{l^2-4\alpha}{2}}$ \\
  $\forall n(r)$ & $\cosh\left(\frac{\rho}{2\,\sqrt{\alpha}}\right)$ & $2\,\beta\,\sqrt{\alpha}\,\sinh\left(\frac{\rho}{2\,\sqrt{\alpha}}\right)$ &
  $\frac{3}{4\alpha}$ & -  \\
  Corrected BTZ & $\cosh\left(\frac{\rho}{2\,\sqrt{\alpha}}\right)$ &
$2\,\beta\,\sqrt{\alpha}\,\sinh\left(\frac{\rho}{2\,\sqrt{\alpha}}\right)$ &
$\frac{3}{4\alpha}$ &
$l^2=4\,\alpha$ \\
  Corrected BTZ & $\pm 1$ & $2\,\beta\,\sqrt{\alpha}\,\sinh\left(\frac{\rho}{2\,\sqrt{\alpha}}\right)$
  & $\frac{1}{4\alpha}$ & $l^2=12\alpha$\\
\hline
\end{tabular}
\end{center}
\caption{BTZ String-Like Solutions in Five-Dimensional Braneworlds
of Codimension-2.}\label{table1}
\end{table}
In this table $l$ is the length of three-dimensional AdS space. The BTZ solution is given by \cite{Banados:1992wn}
\begin{equation}
n^{2}(r)=-M+\frac{r^{2}}{l^{2}}~.\label{btz}
\end{equation}
When its mass is positive, the black hole has a horizon at $r=l\sqrt{M}$, and the radius of
curvature of the $AdS_3$ space $l=(-\Lambda_3)^{-1/2}$ provides the necessary length scale to
define this horizon. For the mass $-1<M<0$, which is dimensionless, the BTZ black hole has a
naked conical singularity, while for $M=-1$ the vacuum $AdS_3$ space is recovered.

The corrected BTZ solution corresponds to a BTZ black hole with
a short distance correction term,
\begin{equation}
n(r)=\sqrt{-M+\frac{r^2}{l^2}-\frac{\zeta}{r}}~,
\label{nsolu5final}
\end{equation}
and it describes a BTZ solution conformally coupled to a scalar
field~\cite{Zanelli1996}.

To introduce a brane we solve the corresponding junction
conditions given by the boundary Einstein equations
(\ref{einsteincomb3}) using the induced metric shown in
(\ref{3dmetric}). For the case when $n(r)$ corresponds to the
BTZ black hole~(\ref{btz}), and the brane
cosmological constant is given by $\Lambda_{3}=-1/l^{2}$, we
found that the energy-momentum tensor in (\ref{einsteincomb3}) is null.
Therefore, the BTZ black hole is localized on the brane in vacuum.

When $n(r)$ is of the form given by (\ref{nsolu5final}), the energy momentum
tensor necessary to sustain such a solution on the brane is given by
\begin{equation}
T_\alpha ^\beta =  \hbox{diag } \left(
\frac{\zeta}{2r^3},\frac{\zeta}{2r^3},-\frac{\zeta}{r^3} \right)\,,
\label{braneEnerMom}
\end{equation}
which is conserved on the brane~\cite{Kofinas:2005a}.

These solutions extend the brane BTZ black hole into the bulk.
The warp function $f^{2}(\rho)$ gives the shape of a 'throat' to the horizon, whose size is defined by the scale $\sqrt{\alpha}$, which is fine-tuned to the length scale of the five-dimensional AdS space.

\section{The Kerr-Schild Method}

In this section we will apply the generalized Kerr-Schild method to the metrics analyzed in~\cite{CuadrosMelgar:2007jx} to generate new solutions. This procedure consists on defining a new metric $\hat{g}_{MN}$ starting from a known one $g_{MN}$ as follows
\begin{equation}\label{newmetric}
\hat{g}_{MN} = g_{MN}\, +\, 2\,H(r,\rho)\, \ell_M\, \ell_N \,,
\end{equation}
and checking if this new metric $\hat{g}_{\mu\,\nu}$ satisfies the Einstein equations. Here $H$ is an arbitrary function of the coordinates $r$ and $\rho$ in this specific case, and  $\ell_{M}$ is a null geodesic vector (in the background metric) described below (\ref{transverse}).

A.~H.~Taub~\cite{taub} introduced this approach in the form just described. There is an extended family of solutions generated by this method. Among others, it reproduces the standard Einstein's solutions: Schwarzschild, Reissner-Nordstr\"om, and (obviously) the Kerr solutions starting from a flat, empty spacetime.
We extend this formalism to the BTZ string-like solutions described in the previous section.

Our case is quite different from those just described since it includes the Gauss-Bonnet term in the original action (\ref{AcGBIG}). The resulting equations are at most quadratic in the function $H(r,\rho)$ due to the index contractions displayed in the GB term.

\subsection{Null Geodesic Vector}

Let us consider a vector $\ell_M = \ell_M (r,\rho)$ which obeys the following conditions,
\begin{eqnarray}
\ell_{M;N}\ell^N =0\,,\quad \ell_M \ell^M =0\,\label{transverse}
\end{eqnarray}
where capital indices run over all the coordinates $(t,r,\phi,\rho,\theta)$.

The definition of the following operator will show to be useful in solving these equations,
\begin{equation}
\hat{D}\,=\,\left[ n^{2}\,\ell_{r}\,\displaystyle\,
\frac{\partial}{\partial\,r}\,+\,f^{2}\,\ell_{\rho}\,\displaystyle\,\frac{\partial}{\partial\,\rho}
\right] .\label{operator}
\end{equation}

Using this operator and the geodesic equation (\ref{transverse}), we obtain the following
equations for the $t$, $\phi$, and $\theta$ components of the null vector $\ell_{A}$,
\begin{equation}\label{li}
\displaystyle\,\frac{1}{f^2}\,\hat{D}\,\ell_{i}\,=\,0\,,\quad\mbox{for i\,=\,t,\,$\phi$ and
$\theta$.}
\end{equation}

The $r$ and $\rho$ components are more involved and read,
\begin{eqnarray}\label{twoeqts}
\displaystyle\,\frac{1}{f^2}\,\hat{D}\, \ell_r + \displaystyle\,\frac{1}{f^2}\,\left[
\displaystyle \frac{\dot{n}}{n^3}\,\ell_t ^2 + n\,\dot{n}\, \ell_r ^2 - \frac{1}{r^3}
\ell_\phi ^2\right] = 0\,,\nonumber
\\
\\
\displaystyle\,\frac{2}{f^2}\,\hat{D}\, \ell_{\rho}\, -\,
\left(\frac{1}{f^2}\right)'\,\left[\frac{1}{ n^2}\, \ell_t ^2 -{n^2}\,\ell_r ^2
-\frac{1}{r^2}\, \ell_\phi ^2\right]\, +\,\left( \frac{1}{b^2}\right)'\, \ell_\theta ^2
=0.\nonumber
\end{eqnarray}
 In our convention over-dot  means differentiating with respect to $r$ while  a prime corresponds to a
derivative with respect to $\rho$.

An additional constraint comes from $\ell_M$ being a null vector (\ref{transverse}). This
condition is

\begin{equation}\label{tracel}
\frac{1}{f(\rho)^2}\,\left[-\frac{\ell_t ^2}{n^2} + {n^2}\,\ell_r ^2 + \frac{\ell_\phi
^2}{r^2}\,\right] + \ell_\rho ^2 + \frac{\ell_\theta ^2}{b^2} = 0\,.
\end{equation}

The Kerr-Schild formalism requires to know at least one explicit solution for the null
geodesic $\ell_{A}$ before we begin our search for a new solution of Einstein's equations. We
start considering the following assumptions about the coordinate dependence of the radial
brane and bulk components of the lightlike vector, $\ell_r$ and  $\ell_{\rho}$

\begin{equation} \ell_{r}\,=\,\ell_{r}(r), \qquad\mbox{and}\qquad
\ell_{\rho}\,=\,\ell_{\rho}(\rho).\label{sep} \end{equation}

With these assumptions a solution for the set of Eqs.(\ref{li}) is readily found. For the
components $i\,=\,t,\,\theta$, and $\phi$ we have

\begin{eqnarray}
\ell_i(r,\rho)\, =\,C_i\,\exp \left( \kappa_1\,\ell_{\rho}\,\int
\frac{dr}{n^2\,\ell_r(r)}\,\right)\times\, \exp \left( -\kappa_1\,\ell_{r} \int \frac{
d\rho}{f^2(\rho)\,\ell_{\rho}}\right).
\end{eqnarray}

A simple non-trivial solution can be obtained setting $\kappa_1\,=\,0$. In this case the
solutions are
\begin{equation}
\ell_t = E, \,\,\ell_{\phi}\, =\, L \,,\,\,\mbox{and}\,\,\, \ell_{\theta}\, =\, K, \,
\end{equation}
where $E$,  $L$, and $K$ are constants related to the energy and the brane and bulk
components of the angular momentum of the particle following the geodesic.

There are still two components left to solve $\ell_{r}(r)$ and $\ell_{\rho}(\rho)$.
Introducing these expressions in the Eqs.(\ref{twoeqts}) and doing the corresponding
simplifications we arrive to

\begin{equation}
n(r)^2\, \dot{\left(\ell_{r}^{2}\right)}\,+\, \left[ -\,
\dot{\left(\frac{1}{n^{2}}\right)}\,E^{2} + \dot{(n^{2})} \, \ell_{r}^{2} +
\dot{\left(\frac{1}{r^{2}}\right)} \, L^{2}\,\right]\, =\, 0. \label{ele1}
\end{equation}

\begin{equation}
\left(\ell_{\rho}^2\right)^{'}\,+\,\left(\displaystyle\frac{1}{f^{2}}\right)^{'}\, \left[ -
\frac{1}{n^{2}} \, E^{2} + n^{2} \, \ell_{r}^{2} + \frac{1}{r^{2}} \, L^{2} \right] \,+ \,
\left(\displaystyle\frac{1}{b^{2}}\right)^{'} \, \ell_{\theta}^{2} = 0 \label{ele3}
\end{equation}

Let us work with the last equation. Employing the null vector condition (\ref{tracel})
and factorizing conveniently we obtain an expression that can easily be integrated
$$
\frac{1}{f(\rho)^2}\,\left[f(\rho)^2\,\left(\ell_{\rho}(\rho)^{2}\,+\,\frac{K^2}{b(\rho)^2}\,\right)
\right]^{'}\,=\,0,$$ from here we find  the general solution for $\ell_{\rho}(\rho)$,
\begin{equation}
\ell_{\rho}(\rho)^{2}\,=\,-\frac{K^{2}}{b(\rho)^2}\,+\,\frac{\xi^{2}}{f(\rho)^2}.\label{lrho}
\end{equation}

 A new constant $\xi$ has been introduced here. If $K\,\neq\,0$, Eq.(\ref{lrho}) is valid as long as its right hand side remains positive and $\rho>0$.

We now solve $\ell_r$ from Eq.(\ref{ele1}). This equation can be written as a total
derivative and after one integration becomes
$$
\left[ -\,\frac{E^2}{n(r)^{2}}\,+\,n(r)^{2}\,\ell_{r}^{2}\,+\,\frac{L^2}{r^2}
\right]\,=\,\chi^2,
$$
where $\chi$ is a constant of integration. However, this constant is not a new parameter
since this component must fit in the null vector restriction Eq.(\ref{tracel}). Inserting this
expression in  Eq.(\ref{tracel}) we obtain that  $\chi^2 \,+\,\xi^2\,=\,0$, fixing $\chi$.

With this last step we have solved Eqs.(\ref{transverse}) for each of the components of
the null geodesic vector.

The null geodesic vector takes the following general expression,

\begin{equation}\label{geovec}
\ell_M = \left(E\, , \,\frac{1}{n^2} \sqrt{E^2 - \left( \frac{L^2}{r^2} +\xi^2 \right)
n^2}\, ,\,L\, ,\,\sqrt{\frac{\xi^2}{f^2} - \frac{K^2}{b^2}} \, , \, K\right)\,.
\end{equation}

\subsection{Solutions Generated by the Kerr-Schild Method}

We choose the following null geodesic vector,
\begin{equation}\label{ngeovec}
\ell_M = \left( 1\,,\, \frac{1}{n^2} \sqrt{1-\frac{L^2}{r^2}n^2}\,,\, L\,,\,0\,,\,0\right)\,,
\end{equation}
where $L$ is a constant associated to the angular momentum of the test particle, and a function $H$ of the form $H(r,\rho)=h_1(r)\,h_2(\rho)$. We use this Ansatz in (\ref{newmetric}) 
and replace back into the Einstein equations obtaining eight non-null equations. Factorizing $(\rho t)$ and $(\rho r)$ components we arrive to the following equation
\begin{equation}\label{famosita}
[2h_{2}(\rho)f'-fh_2 '(\rho)][h_1(r)+h_1 '(r)r]=0 \,.
\end{equation}

Once we solve Eq.(\ref{famosita}), the rest of the equations are automatically fulfilled provided that the corresponding constraints in Table \ref{table1} are satisfied.
   
If we choose to solve this equation for $h_1(r)$, we obtain $h_1(r)=\sigma/r$, where $\sigma$ is a constant, and $h_2(r)$ remains free. When putting back into (\ref{newmetric}) the new metric turns into 
\begin{eqnarray}\label{met}
 ds^2 &=& \left(-f^2n^2+2\frac{\sigma}{r}h_2\right)dt^2-4\frac{\sigma}{r}\frac{h_2}{n^2}dtdr+\left(\frac{f^2}{n^2}+2\frac{\sigma}{r}\frac{h_2}{n^4}\right)dr^2\nonumber \\
&&+f^2r^2d\phi^2+d\rho^2+b^2d\theta^2  \,.
\end{eqnarray}
This metric represents a 5-dimensional gravity solution, which can be diagonalized in few cases depending on the form of $h_2(\rho)$. Notice that the line element (\ref{met}) does not include a boundary membrane.
As our main interest here is to find braneworld solutions, naturally our next step is to embed a brane in this bulk metric. In order to proceed to this point, the first requirement we find is that the induced metric must fulfill the junction conditions, {\it i.e.}, the 3-dimensional Einstein equations (\ref{einsteincomb3}). As far as we know, the BTZ metric is the only solution describing a black hole in a (2+1) spacetime, thus,  
$h_2(\rho)$ becomes constrained to a multiple of the warp factor $f^2(\rho)$ to be able to recover a BTZ-like metric on the brane. In particular, if $h_2(\rho)=f^2(\rho)$, we can make a coordinate transformation to end up with the BTZ string coupled to a brane scalar field as it will be shown below.

Alternatively, if we solve Eq.(\ref{famosita}) for $h_2(\rho)$, we get $h_2(\rho)=f^2(\rho)$, and $h_1(r)$ becomes arbitrary. In this case the choices for $h_1(r)$ are several and give place to the following new solutions. Nevertheless, as the presence of the brane imposes junction conditions (\ref{einsteincomb3}), we should point out that the right options for $h_1(r)$ will be determined by the requirement of yielding a physically meaningful brane energy-momentum tensor.

\subsubsection{Charged BTZ string}

Let us first consider the solution $f(\rho)=\cosh(\rho/2\sqrt\alpha)$ and $b(\rho)=2\beta\sqrt\alpha \sinh(\rho/2\sqrt\alpha)$. 
By choosing $h_1(r)=\frac{Q^2}{2} \ln\, r$ and $L=0$ we arrive to the following metric,
\begin{eqnarray}\label{metq}
ds^2 = f^2 \left[-(n^2-Q^2 \ln\, r)\,dt^2 + \frac{2Q^2 \ln\, r}{n^2} dt\,dr + \left(\frac{n^2+Q^2 \ln\, r}{n^4}\right) dr^2 + r^2 d\phi^2 \right]\nonumber \\
+ d\rho^2 + b^2 d\theta ^2\,. \qquad \qquad \qquad \qquad \qquad \qquad  \qquad \qquad \qquad \qquad \qquad \qquad \quad
\end{eqnarray}
In order to obtain a more familiar form of the metric we make the following coordinate transformation,
\begin{equation}
dt = d\hat t + \frac{Q^2 \ln\, r\, dr}{n^2 (n^2-Q^2\ln \, r)}\,.
\end{equation}
This change cancels out the non-diagonal term such that the metric (\ref{metq}) turns into
\begin{equation}\label{charge}
ds^2 = f^2 \left(-\hat n^2 d\hat t^2 + \frac{dr^2}{\hat n^2} + r^2 d\phi^2\right) + d\rho^2 + b^2 d\theta ^2\,,
\end{equation}
with $\hat n^2 = -M+r^2/l^2 -Q^2 \ln\, r$, which describes a charged BTZ string whose charge is confined to the brane. In order to verify this statement we calculate the brane energy momentum tensor necessary to hold this solution and we obtain
\begin{equation}
T_\mu ^\nu = diag \left(-\frac{Q^2}{2r^2}\,,\,-\frac{Q^2}{2r^2}\,,\,\frac{Q^2}{2r^2} \right)\,.
\end{equation}
This is precisely the stress energy tensor related to a charged object in (2+1) dimensions.

\subsubsection{BTZ string coupled to a brane scalar field}

Working with the same expressions for $f(\rho)$ and $b(\rho)$ 
we now choose $h_1(r)=\frac{\zeta}{2r}$ and the resulting metric is
\begin{eqnarray}\label{mets}
ds^2 &=& f^2 \left[-\left(n^2-\frac{\zeta}{r}\right)\,dt^2 + \frac{2\zeta}{n^2 r}
dt\,dr + \left(\frac{b^2r^2+\zeta r}{n^4 r^2}\right) dr^2 + r^2 d\phi^2 \right]\nonumber \\
&&+ d\rho^2 + b^2 d\theta ^2\,. \qquad \qquad \qquad \qquad \qquad \qquad  \qquad
\qquad \qquad \qquad \qquad \qquad \quad
\end{eqnarray}
We now make the coordinate transformation
\begin{equation}
dt = d\hat t + \frac{\zeta}{n^2 r (n^2-\zeta/r)}\,,
\end{equation}
to arrive to the following metric
\begin{equation}\label{scalar}
ds^2 = f^2 \left(-\hat n^2 d\hat t^2 + \frac{dr^2}{\hat n^2} + r^2 d\phi^2\right) + d\rho^2 + b^2 d\theta ^2\,,
\end{equation}
where $\hat n^2 =  -M+r^2/l^2 - \zeta/r$, which corresponds to a BTZ black hole coupled to a scalar field on the brane. This solution had already been found in~\cite{CuadrosMelgar:2007jx}.

\subsubsection{Charged BTZ string coupled to a brane scalar field}

Another possible combination is to choose the original metric to be (\ref{charge}) and $h_1(r)=\zeta/2r$. Following the same procedure as the previous case, we find the same metric as (\ref{scalar}), but with $\hat n^2 = -M+r^2/l^2 -Q^2 \ln\, r - \zeta/r$. If we compute the energy-momentum tensor on the brane, we find
\begin{equation}
T_\mu ^\nu = diag \left(-\frac{Q^2}{2r^2} + \frac{\zeta}{2r^3} \,,\,-\frac{Q^2}{2r^2}+ \frac{\zeta}{2r^3}\,,\,\frac{Q^2}{2r^2}- \frac{\zeta}{r^3} \right)\,,
\end{equation}
which corresponds to a charged BTZ black hole coupled to a scalar field on the brane.

\subsubsection{BTZ string with angular momentum}

In order to add angular momentum to the original BTZ string, we pick $h_1(r)=c$, where $c$ is a constant. In this case $L\not= 0$. Thus, the metric takes the following form,
\begin{eqnarray}
ds^2 &=&  f^2 \left[-(n^2-2c) dt^2 + \frac{4c}{n^2} \sqrt{1-\frac{L^2}{r^2}n^2}\, dt\,dr + 4cL \,dt\,d\phi \right. \nonumber\\
 &&+\left.\frac{n^2 r^2 + 2cr^2 -2cn^2 L^2}{n^4 r^2} dr^2 + \frac{4cL}{n^2} \sqrt{1-\frac{L^2}{r^2}n^2} dr\,d\phi \right.\nonumber \\
&&+ \left.(r^2 +2cL^2) d\phi ^2 \right] + d\rho^2 +b^2 d\theta^2\,.
\end{eqnarray}
Introducing the following transformations,
\begin{eqnarray}
dt &=& d\hat t + u(r) dr \\
d\phi &=& d\hat\phi + v(r) dr \,,
\end{eqnarray}
with
\begin{eqnarray}
u(r) &=& \frac{2cr^2}{n^2(n^2 r^2 -2r^2c + 2cL^2n^2)} \sqrt{1-\frac{L^2}{r^2}
n^2} \\
v(r) &=& -L \frac{n^2}{r^2} u(r)\,.
\end{eqnarray}
Moreover, if we define $R^2 = r^2 + 2cL^2$, $J=-4L$, and $\tilde M=M + 2c(L^2/l^2 + 1)$, we arrive to the metric for a rotating BTZ-string,
\begin{equation}
ds^2 = f^2 \left[-\hat n^2 d\hat t^2 + \frac{dR^2}{\hat n^2} + R^2 \left(\frac{-J}{2R^2} d\hat t + d\hat\phi\right)^2 \right] +d\rho^2 +b^2 d\theta^2\,,
\end{equation}
where $\hat n^2(r) = -\tilde M + R^2/l^2 + J^2/4R^2$. The corresponding energy-momentum tensor on the brane can be calculated, and we find that it vanishes, as it should be for a rotating BTZ brane black hole.

\bigskip

Analogously, when we use the solutions $f(\rho)=1$, $b(\rho)=\gamma \sinh (\rho/\gamma)$, and $f(\rho)=1$, $b(\rho)=2\beta \sqrt\alpha \sinh(\rho/2\sqrt\alpha)$, we also arrive to several solutions involving angular momentum and scalar fields. We should stress that in these cases we did not find any charged solution.

\section{Geodesic Structure}

In this section we study the geodesic behaviour in the background of the solutions displayed in Table \ref{table1}.

Let us begin our study by considering the Lagrangian for the BTZ black hole in codimension-2 branes with the solution $f(\rho)=\cosh\left(\frac{\rho}{2\sqrt{\alpha}}\right)$ and $b(\rho)=2\beta\sqrt\alpha\sinh\left(\frac{\rho}{2\sqrt\alpha}\right)$,
\begin{equation}\label{lag}
{\cal L} = \cosh\left(\frac{\rho}{2\sqrt\alpha}\right)^{2}\Bigg(-n(r)^{2}\dot{t}^{2}+\frac{\dot{r}^{2}}{n(r)^{2}}+r^{2}\dot{\phi}^{2}\Bigg)+\dot{\rho}^{2}+4\beta^{2}\alpha\sinh\left(\frac{\rho}{2\sqrt\alpha}\right)^{2}\dot{\theta}^{2} \,,
\end{equation}
where a dot indicates derivative with respect to the affine parameter $\lambda$.

As it is independent of $t$, $\theta$ and $\phi$, we can write the following equations of motion,
\begin{eqnarray}
\frac{\partial {\cal L}}{\partial\dot{t}}&=&-2\cosh\left(\frac{\rho}{2\sqrt\alpha}\right)^{2}n(r)^{2}\dot{t}=-2E\,, \label{affinet}\\
\frac{\partial {\cal L}}{\partial\dot{\phi}}&=&2\cosh\left(\frac{\rho}{2\sqrt\alpha}\right)^{2}r^{2}\dot{\phi}=2L\,, \\
\frac{\partial {\cal L}}{\partial\dot{\theta}}&=&8\beta^{2}\alpha\sinh\left(\frac{\rho}{2\sqrt\alpha}\right)^{2}\dot{\theta}=2K\,,
\end{eqnarray}
where $L$ and $K$ are the angular momenta of the particle related to $\phi$ and $\theta$ coordinates, respectively. Notice that Eq.(\ref{affinet}) gives us a relation between the coordinate time $t$ and the affine parameter $\lambda$.

With these equations the Lagrangian (\ref{lag}) can be written as
\begin{eqnarray}\label{lag2}
{\cal L}&=&\cosh^2\left(\frac{\rho}{2\sqrt\alpha}\right)\left[\frac{-E^2}{\cosh^4\left(\frac{\rho}{2\sqrt\alpha}\right) n(r)^{2}}+\frac{\dot{r}^{2}}{n(r)^{2}}+
\frac{L^{2}}{r^{2}\cosh^4\left(\frac{\rho}{2\sqrt\alpha}\right)}\right] \nonumber \\
&&+\dot{\rho}^{2}+\frac{K^{2}}{4\beta^{2}\alpha\sinh^2\left(\frac{\rho}{2\sqrt\alpha}\right)}=h\,,
\end{eqnarray}
where $h=0$ or $-1$ is a parameter describing lightlike and timelike geodesics, respectively.

\subsection{Geodesics on the Brane}

Let us consider a particle with $K=0$. Using Eq.(\ref{lag2}) at the position of the brane ($\rho=0$) we can find the effective potential for the geodesic motion,
\begin{equation}\label{potential}
\dot{r}^{2}=E^{2}-n^2(r)\left(\frac{L^{2}}{r^{2}}-h\right) \quad \Rightarrow \quad V_{eff}^{2}=n^2(r)\left(\frac{L^{2}}{r^{2}}-h\right)\,.
\end{equation}

This equation can be integrated to obtain the orbits as well,
\begin{equation}\label{orbit}
\frac{dr}{d\lambda}=\sqrt{E^{2}-n^2(r)\left(\frac{L^{2}}{r^{2}}-h\right)}\,.
\end{equation}

\subsubsection{BTZ Case}

For radial geodesics ($L=0$) the effective potential becomes
\begin{equation}
V_{eff}^{2}=n^2(r)\,h=\left(-M+\frac{r^{2}}{l^2}\right)h\,.
\end{equation}
In this case lightlike geodesics are just straight lines. 

\begin{figure}[h!]
\centering
\includegraphics[width=5.5cm]{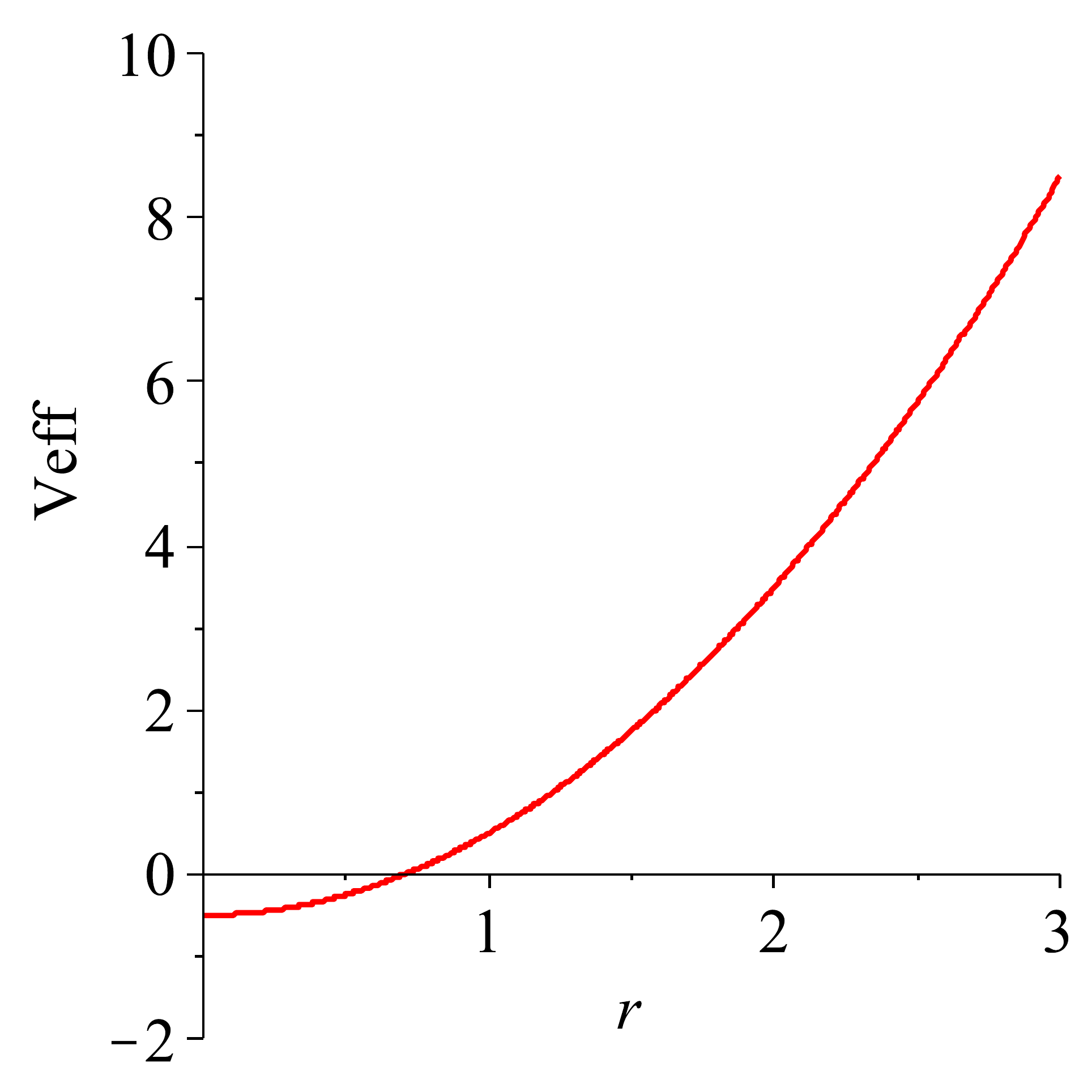}
\includegraphics[width=6.5cm]{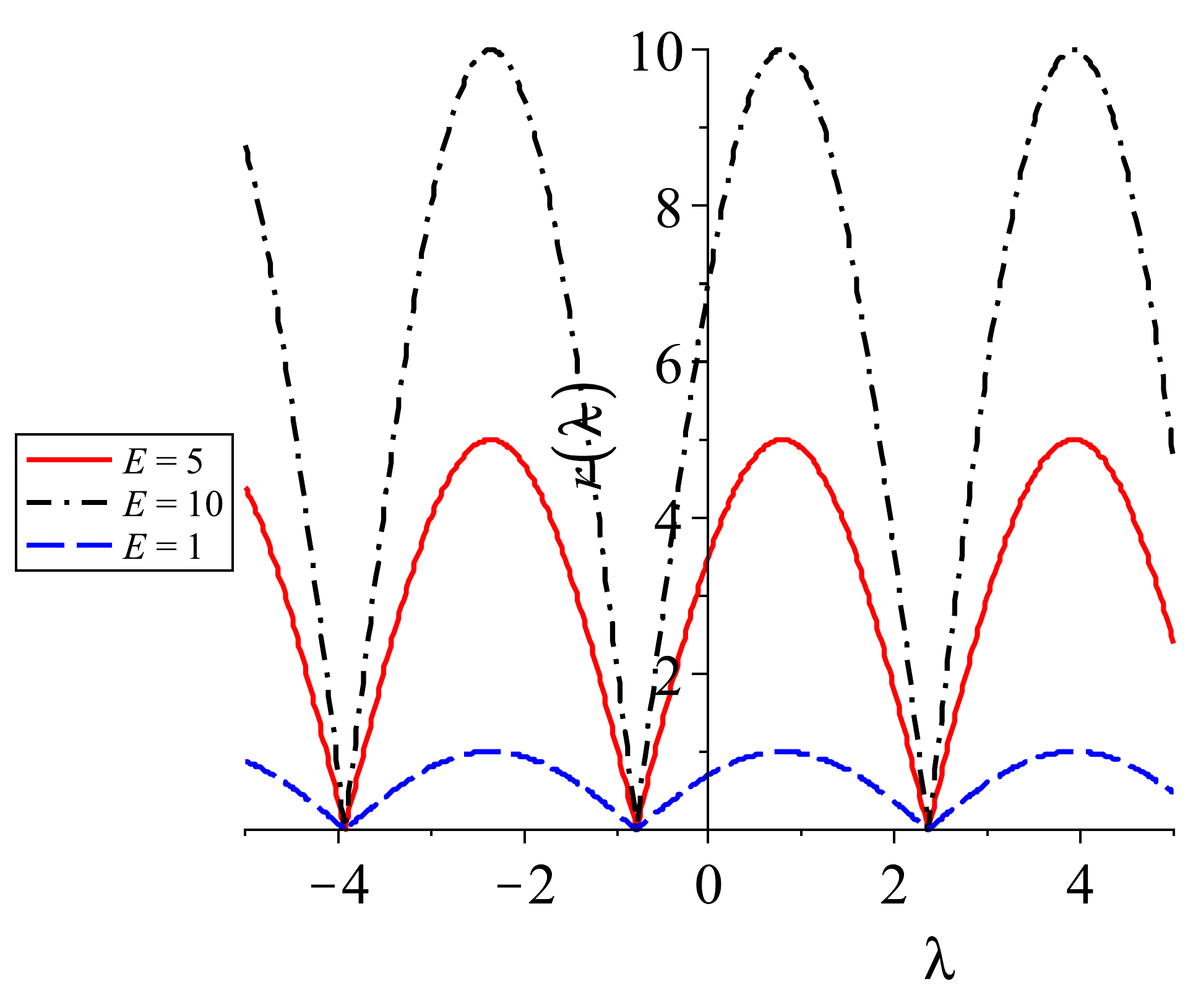}
\caption{Effective potential (left) and orbits (right) for radial timelike particles on the brane. These graphs correspond to $l=1$, $M=0,5$. Notice that geodesics have an oscillatory behaviour outside the event horizon, however, there are no stable orbits since some of the particles could cross it and never return. Particles with $E<0$ are not allowed since $V_{eff}=0$ at the event horizon.}
\label{Fig.1}
\end{figure}

For the timelike case, we display the potential in Fig.\ref{Fig.1}.
The corresponding orbits can be obtained by direct integration of Eq.(\ref{orbit}),
\begin{equation}
r(\lambda) = l\sqrt\frac{E^2+M}{2}\left[1+\sin(2\lambda)\right]^{1/2}\,,
\end{equation}
and they are shown in Fig.\ref{Fig.1}. We see that geodesics have an oscillatory behaviour outside the event horizon, however, there are no stable orbits since some of the particles could cross it and never return. Notice that particles with $E<0$ are not allowed since $V_{eff}=0$ at the event horizon.

\bigskip

For particles with angular momentum the effective potential turns out to be
\begin{equation}
 V_{eff}^{2}=\left(-M+\frac{r^{2}}{l^{2}}\right)\left(\frac{L^{2}}{r^{2}}-h\right)\,.
\label{veff}
\end{equation}

\begin{figure}[h!]
\centering
\includegraphics[width=6.5cm]{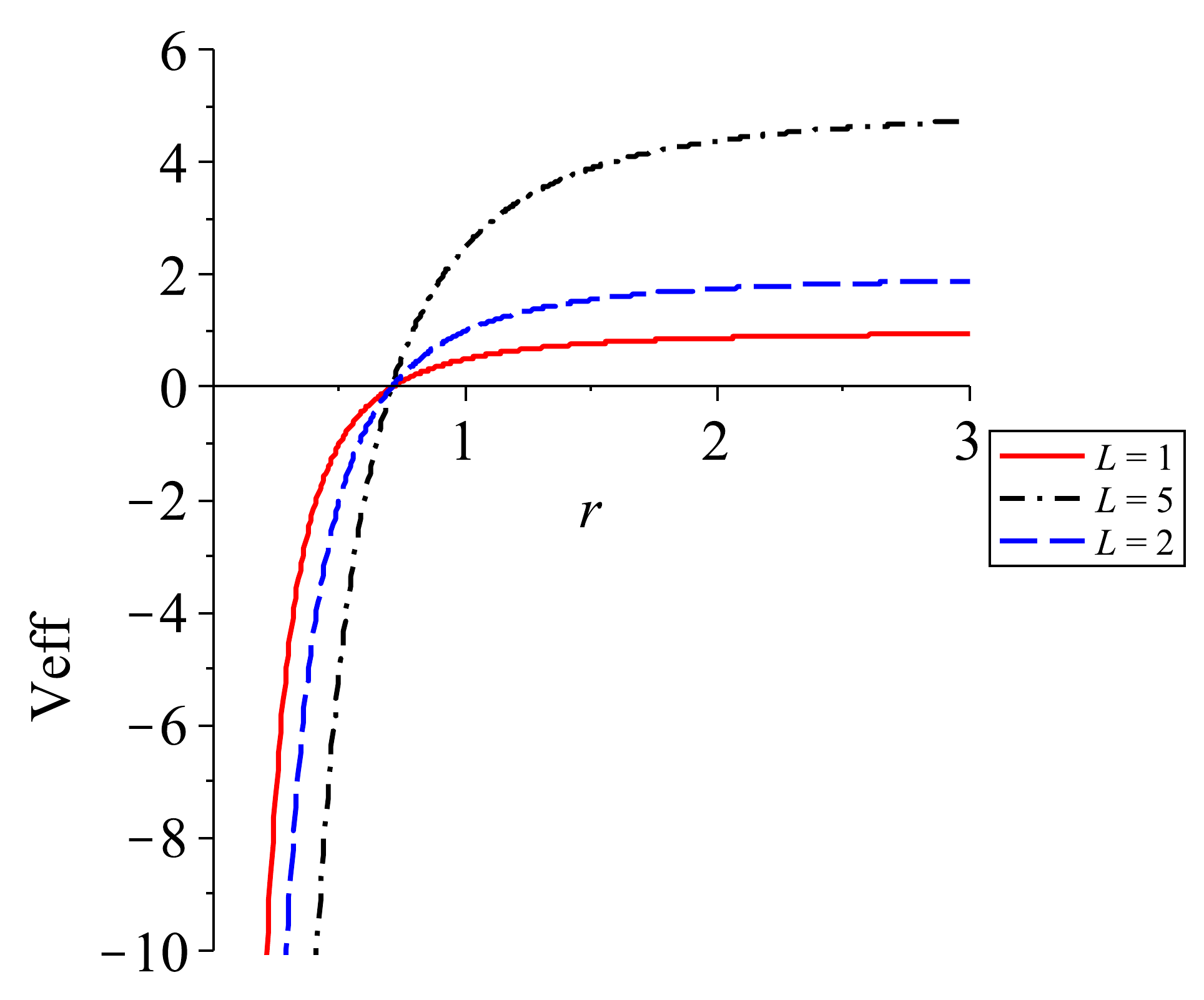}
\includegraphics[width=6.3cm]{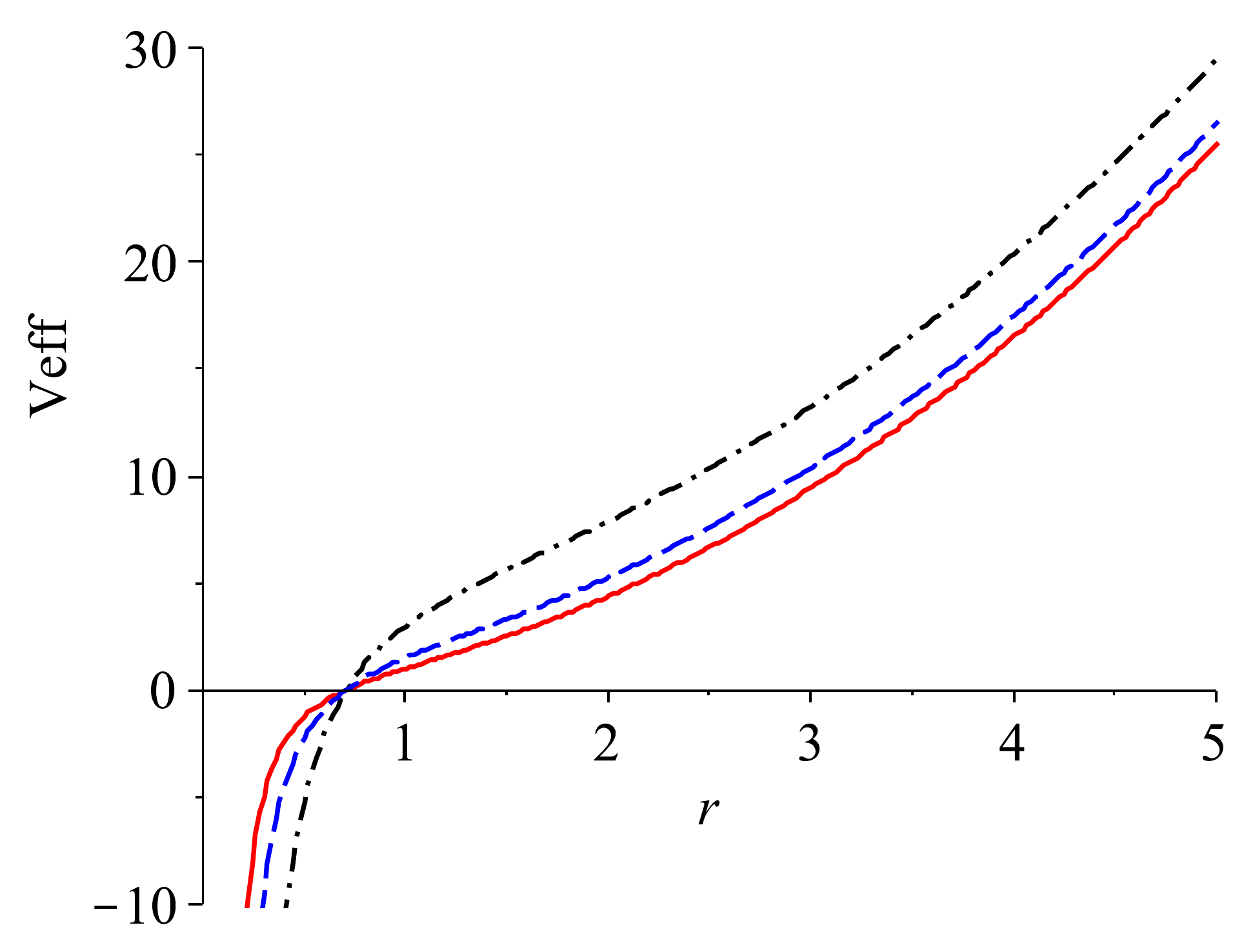}
\caption{Effective potential for lightlike (left) and timelike (right) brane particles with different angular momenta ($L=1,\, 2,\, 5$). Remark that all the potentials cross themselves at the event horizon. 
The lightlike potential has an asymptote given by $L/l$, while the timelike potential grows with no limit.}
\label{Fig.3}
\end{figure}

We plot this potential for lightlike and timelike cases in Fig.\ref{Fig.3}, where we set $M=0,5$, $l=1$, $L=2$, and $L=5$. In both cases the potentials intercept themselves at the event horizon. Remark that far from the event horizon the lightlike potential has an asymptotic behaviour that can be inferred from Eq.(\ref{veff}), {\it i.e.}, $V_{eff} \rightarrow L/l$.


In order to obtain the orbits for the lightlike case ($h=0$) with angular momentum  we can integrate Eq.(\ref{orbit}) to attain,
\begin{equation}
r(\lambda)=\pm\sqrt{\left(E^2-\frac{L^2}{l^2}\right) \lambda^2 - \frac{ML^2}{E^2-L^2/l^2}}\,,
\end{equation}
and we should stress that only the plus sign has physical meaning.

\begin{figure}[h!]
\centering
\includegraphics[width=6.5cm]{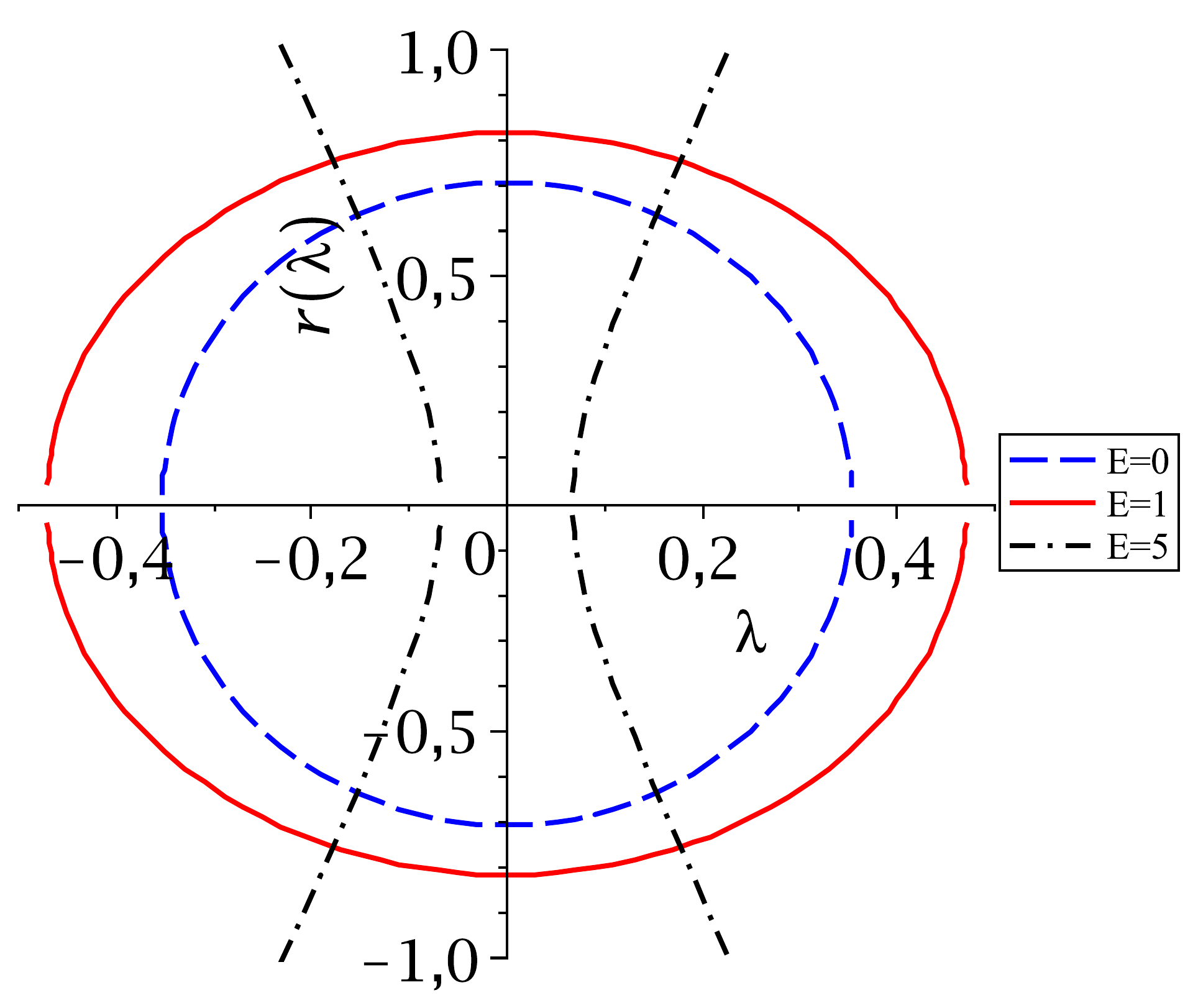}
\includegraphics[width=6.5cm]{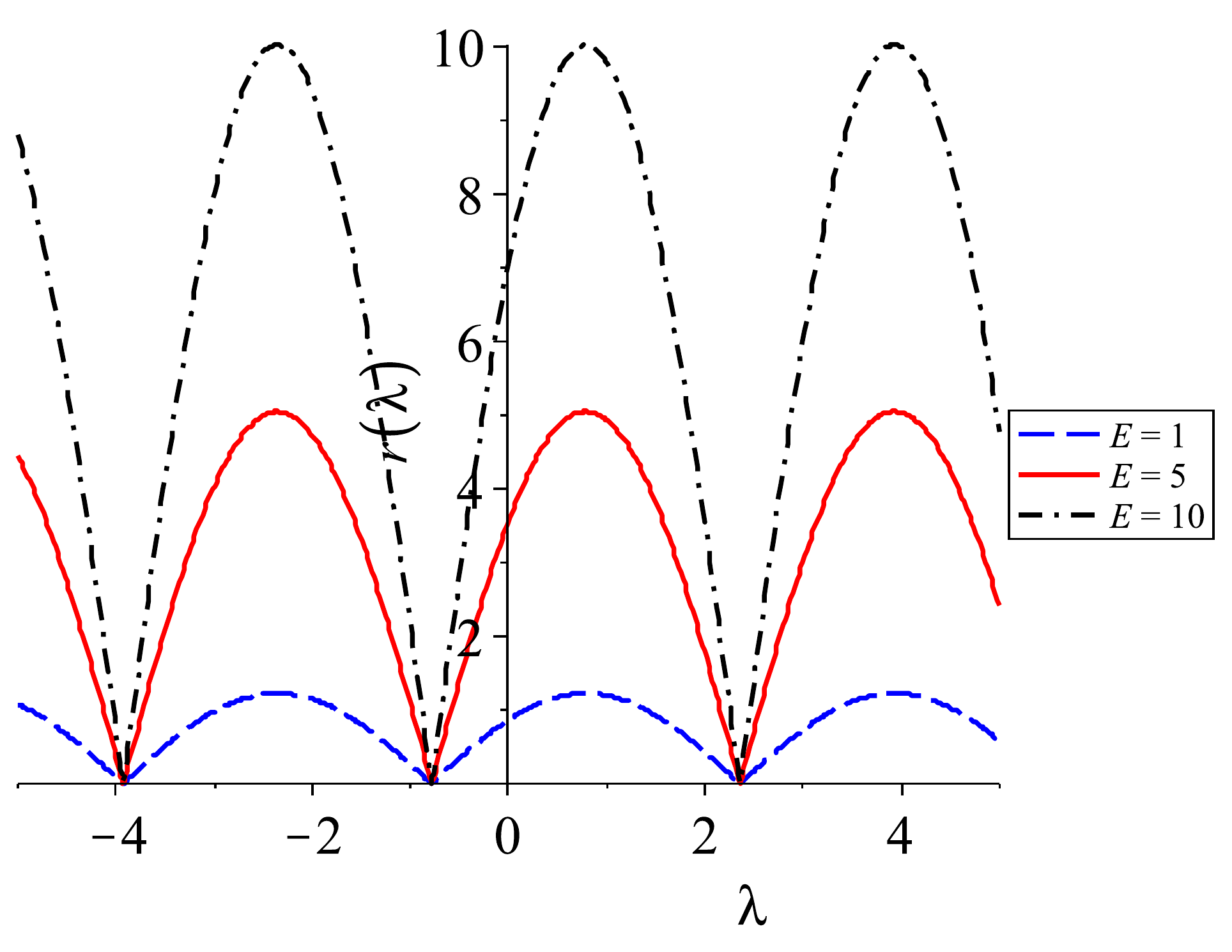}
\caption{Orbits for lightlike (left) and timelike (right) brane geodesics with angular momentum ($L=2$). The energies of the particles are shown in the legend. Notice that in the lightlike case orbits with low energy fall into the event horizon whereas particles with high energy can escape from the black hole. For timelike geodesics the orbits have basically the same shape as those in the radial case (Fig.\ref{Fig.1}b), the only effect of $L$ is to increase the amplitude of the oscillation.}
\label{Fig.4}
\end{figure}

Analogously, we perform the integration of Eq.(\ref{orbit}) in the timelike case and we arrive to
\begin{equation}
r(\lambda) = \sqrt\frac{E^2 l^2+Ml^2-L^2}{2} \left[1+\sqrt{1+\frac{4ML^2}{(E^2+M-L^2/l^2)^2l^2}} \sin(2\lambda)\right]^{1/2}\,.
\end{equation}
The corresponding orbits are presented in Fig.\ref{Fig.4}. From this figure we can see that lightlike geodesics with low energies fall unavoidably into the event horizon. However, particles with high energies can escape from the black hole showing that energy extraction is possible but only with massless particles. On the other side, the timelike orbits have basically the same shape as those in the radial case (Fig.\ref{Fig.1}b), the only effect of $L$ is to increase the amplitude of the oscillation. Again some of them can cross the event horizon depending on the energy of the oscillation. In both lightlike and timelike cases a similar qualitative behaviour was also found in the study of pure BTZ geodesic structure~\cite{cmp}.

\subsubsection{BTZ with Electric Charge and Scalar Field}

\begin{figure}[h!]
\centering
\includegraphics[width=6.7cm]{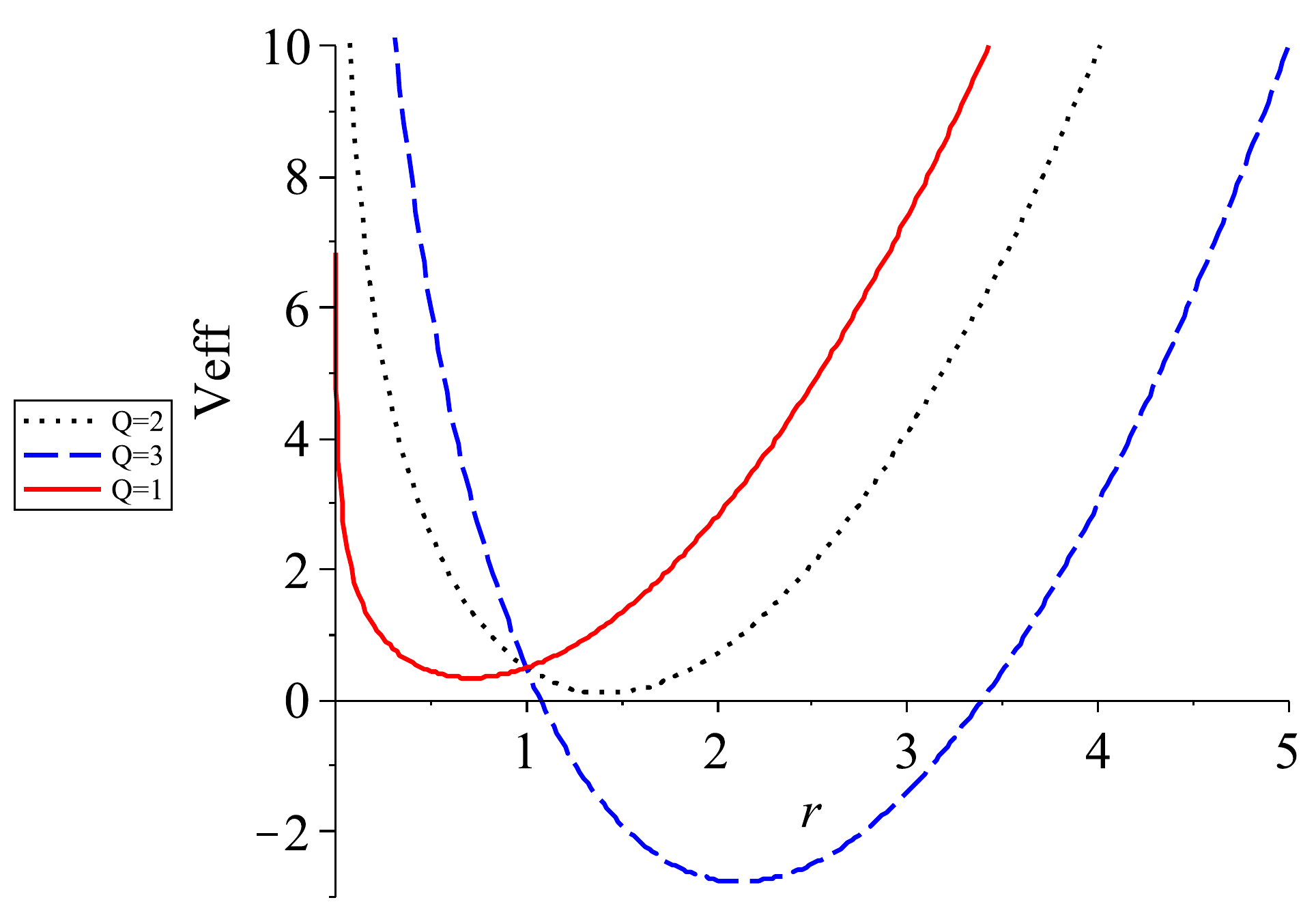}
\includegraphics[width=6.7cm]{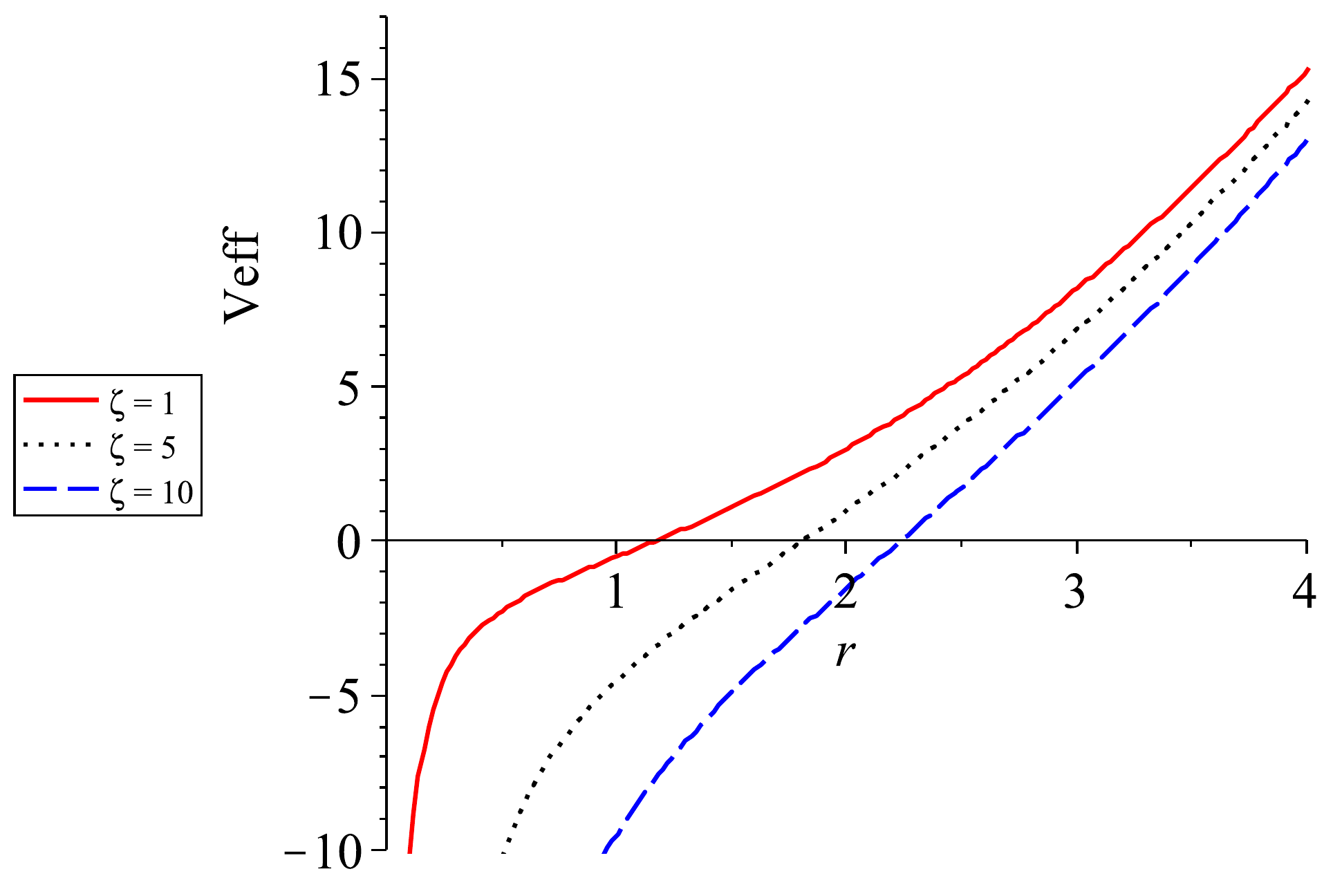}
\caption{Effective potential for timelike geodesics on the brane for BTZ with charge (left) and with scalar field (right). These graphs correspond to $Q=1,2,3$ and $\zeta=1,5,10$. In the former case the potential admits oscillating orbits that can fall into the event horizon when $Q$ is small, however, as $Q$ grows, the minimum of the potential is shifted outside the horizon making possible the existence of stable oscillating or bounded geodesics. In the latter case only unstable oscillations are allowed.}
\label{Fig.9}
\end{figure}

In this case the resulting potential when adding charge or a scalar field to the BTZ solution is again given by Eq.(\ref{potential}) with the corresponding $n(r)$ as follows,
\begin{eqnarray}
n(r)&=&\sqrt{-M+\frac{r^{2}}{l^{2}}-Q^{2}\ln(r)}\,, \qquad \mbox{charged BTZ} \\
n(r)&=&\sqrt{-M+\frac{r^{2}}{l^{2}}-\frac{\zeta}{r}}\,, \qquad \mbox{BTZ + scalar field}
\end{eqnarray}
Both potentials are shown in Fig \ref{Fig.9}.

We can deduce from these graphics that the charged BTZ potential admits oscillating orbits that can fall into the event horizon when $Q$ is small, however, as $Q$ grows, we see that the minimum of the potential is shifted outside the horizon making possible the existence of stable oscillating or bounded geodesics.
 The potential corresponding to BTZ coupled to a scalar field shows that just unstable oscillating orbits are allowed.

\subsection{Geodesics in the Bulk}

Here we study the geodesics that explore the extra dimensions. Although standard particles are not allowed to travel outside the brane, we perform this analysis as a way to acquire a better understanding of the geometry of these solutions. For this analysis we will consider the $r$ coordinate lying outside the black hole horizon, for instance at $r=2\sqrt{M}l=2r_H$. In this case it will be convenient to write Eq.(\ref{lag2}) in a different way,
\begin{equation}\label{dotrho}
\dot{\rho}^{2}=h+\frac{E^{2}}{M\cosh^{2}\left(\frac{\rho}{2\sqrt{\alpha}}\right)}-\frac{L^2}{2Ml^{2}\cosh^{2}\left(\frac{\rho}{2\sqrt{\alpha}}\right)}-\frac{K^{2}}{4\beta^{2}\alpha\sinh^{2}\left(\frac{\rho}{2\sqrt{\alpha}}\right)} \,.
\end{equation}
Defining a new variable $u$ as
\begin{equation}\label{u}
u\,=\,2\,\sqrt{\alpha}\sinh\left(\frac{\rho}{2\sqrt{\alpha}}\right)\,,
\end{equation}
and replacing in Eq.(\ref{dotrho}) it becomes
\begin{equation}\label{dotu}
\dot{u}^{2}= \varepsilon^2-\left[\frac{L^2}{2Ml^2}+\left(1+\frac{u^2}{4\alpha}\right)\left(\frac{K^2}{\beta^2 u^2}-h \right)\right]\,,
\end{equation}
where $\varepsilon^2=E^2/M$. Thus, we can define an effective potential given by
\begin{equation}
V_{eff}^2(u)=\left[\frac{L^2}{2Ml^2}+\left(1+\frac{u^2}{4\alpha}\right)\left(\frac{K^2}{\beta^2 u^2}-h \right)\right]\,.
\end{equation}

\begin{figure}[h!]
\centering
\includegraphics[width=6cm]{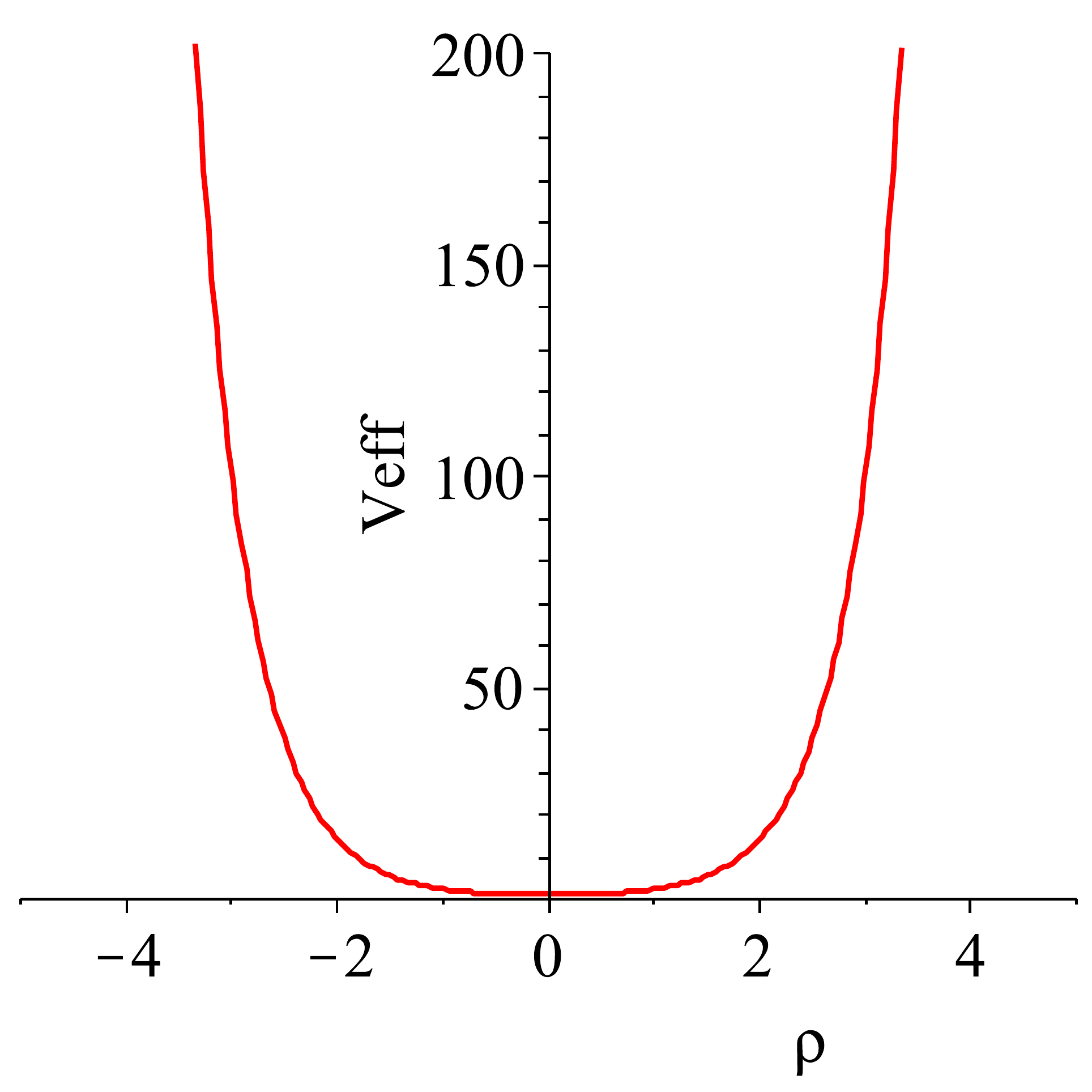}
\includegraphics[width=7cm]{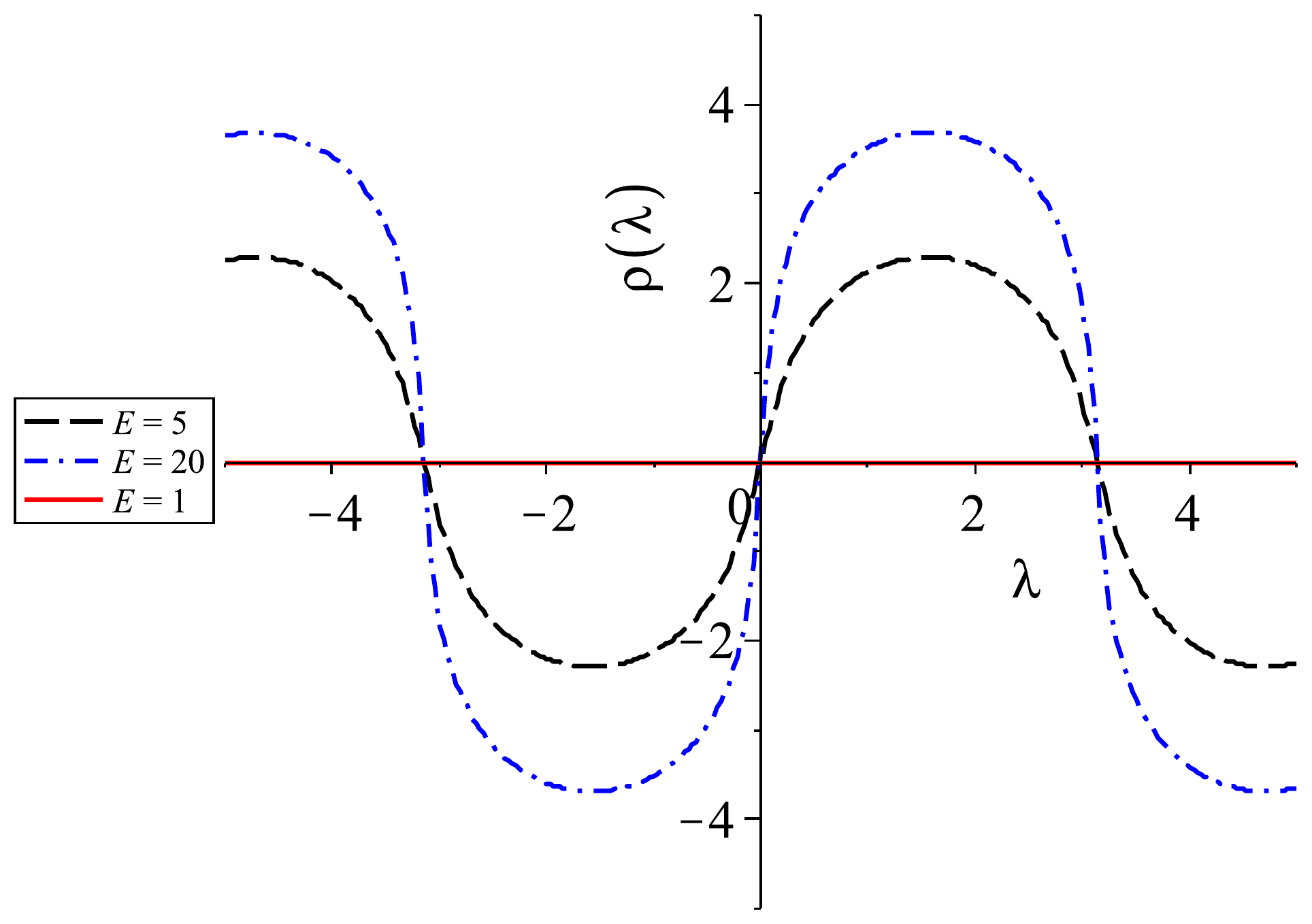}
\caption{Effective potential and timelike orbits for radial particles in the bulk. From the oscillating behaviour of the trajectories we can infer the existence of shortcuts since particles leaving the brane can return to it. Note that the trajectory with the energy that corresponds to the minimum of the effective potential ($E=1$) is entirely on the brane.}
\label{Fig.6}
\end{figure}

For radial geodesics ($L=K=0$) notice that as the effective potential vanishes in the lightlike case, the orbits are just straight lines. On the other side, the orbits for timelike geodesics can be found by integrating Eq.(\ref{dotu}) and replacing $u$ from Eq.(\ref{u}). Thus, we obtain,
\begin{equation}
\rho(\lambda)= 2\sqrt\alpha\; \hbox{arcsinh}\left[\sqrt{E^2-1}\,\sin\left(\frac{\lambda}{2\sqrt\alpha}\right)\right]\,.
\end{equation}
Some of the orbits are depicted in Fig.\ref{Fig.6}. 
This graph shows oscillating trajectories that cross the brane. This implies that particles leaving the brane can return in the future. This fact opens up the possibility for the existence of shortcuts, paths connecting two points which are shorter in the bulk than on the brane~\cite{sc}. In addition, note that the trajectory with the energy that corresponds to the minimum of the effective potential is on the brane.

Now we turn to the case of timelike and lightlike geodesics with $K\ne 0$ and/or $L\ne 0$.

In the lightlike case, if $K=0$ and $L \ne 0$ we have a constant potential $V_{eff}=\frac{L^2}{2Ml^2}$. The corresponding timelike case has the same potential as the $L=0$ geodesics, only shifted by a constant, thus, the shape of the geodesics is the same as the one shown in  Fig.\ref{Fig.6}b. 

\begin{figure}[h!]
\centering
\includegraphics[width=6.5cm]{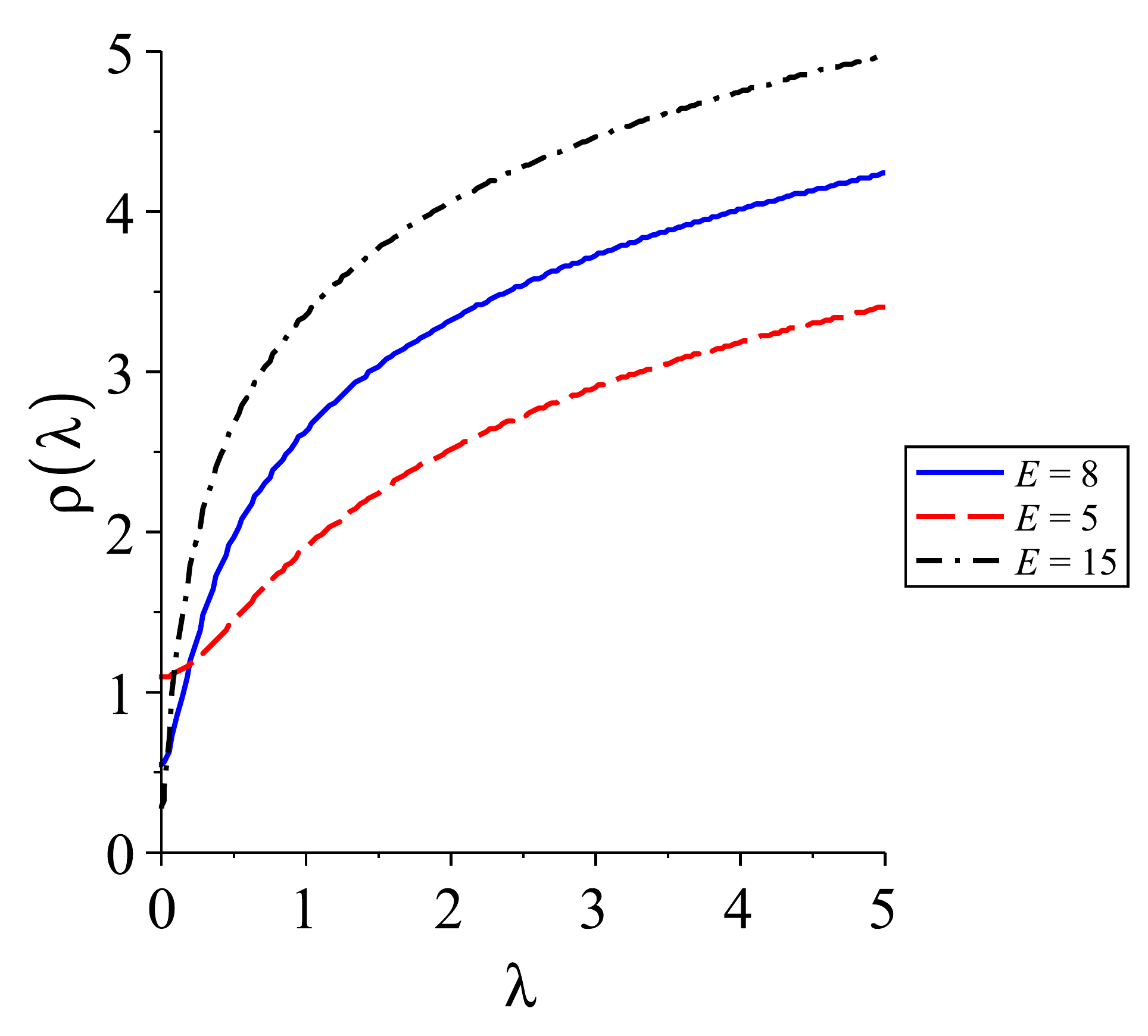}
\includegraphics[width=6.3cm]{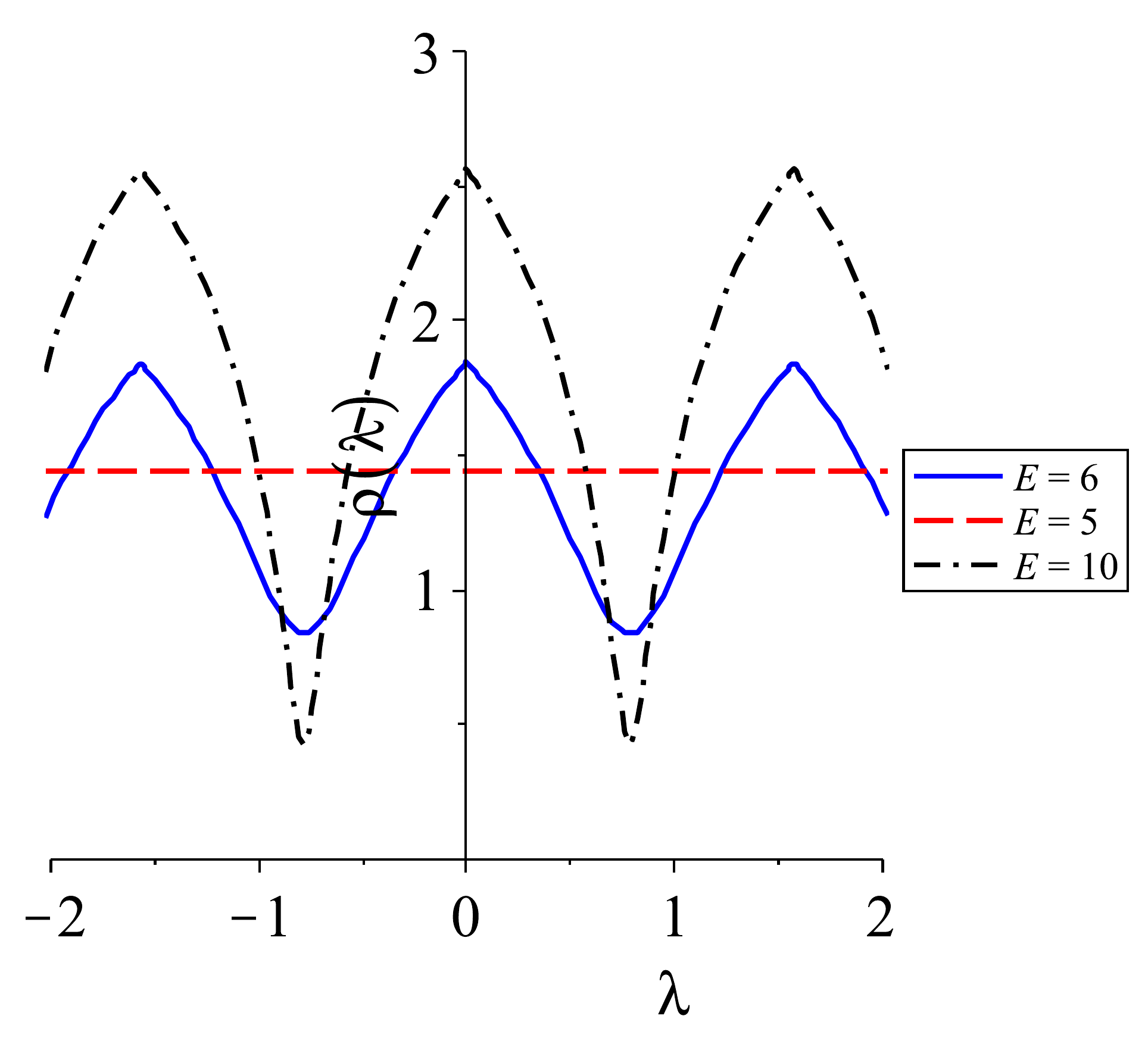}
\caption{Orbits for lightlike (left) and timelike (right) bulk particles with angular momentum ($K=2$, $L=0$). Lightlike geodesics show a scattering behaviour. While timelike particles oscillate near the brane, and the higher their energies are, the closer to the brane they can reach, but never cross it.}
 \label{Fig.7}
\end{figure}

When $K \not= 0$ and $L=0$, we obtain the orbits displayed in Fig.\ref{Fig.7}.
We can notice that in the lightlike case the particles seem to be scattered by a barrier-like potential near the brane. Regarding the timelike geodesics, we can see an oscillating behaviour around certain position parallel to the brane, which corresponds to the orbit with the minimal permitted energy. Additionally, the higher the energies of the particles are, the closer to the brane they can reach, but never cross it.

\bigskip

\begin{figure}[h!]
\centering
\includegraphics[width=7cm]{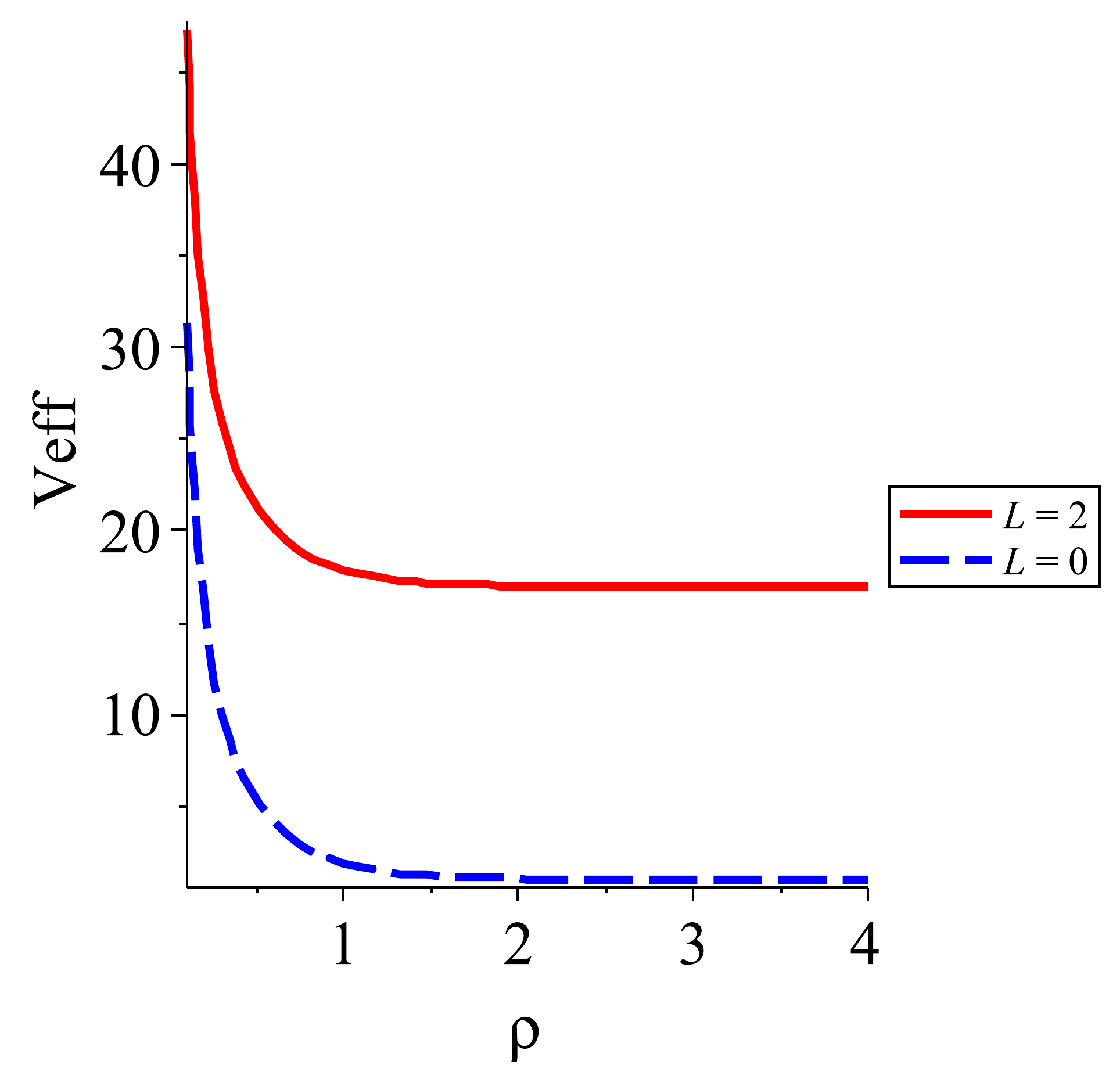}
\caption{Bulk effective potential for $f=1$, with $L=0,2$ and $K=1$. We can see that close to the brane the potential has a scattering behaviour, while far from the brane it becomes constant, so that any bulk geodesic behaves like a free particle.}
\label{Fig.8}
\end{figure}

Forthwith, let us check the geodesic behaviour for the solution $f(\rho)=1$ and $b(\rho)=\gamma\sinh(\rho/\gamma)$ (see Table {\ref{table1}}). By fixing $r$ we can write an analogous equation to (\ref{lag2}),
\begin{equation}
{\cal L}=-\frac{E^{2}}{n^{2}(r)}+\frac{L^{2}}{r^{2}}+\dot{\rho}^{2}+\frac{K^{2}}{\gamma^{2}\sinh^2\left(\rho/\gamma\right)}=h\,.
\end{equation}
So that,
\begin{equation}
 \dot{\rho}^{2}=\frac{E^{2}}{n^{2}(r)}-\frac{L^{2}}{r^{2}}-\frac{K^{2}}{\gamma^{2}\sinh^2\left(\rho/\gamma\right)}+h\,.
\end{equation}
Fixing $r$ the effective potential becomes,
\begin{equation}
V^2 _{eff}= \frac{L^{2}}{2Ml^{2}}+\frac{K^2}{\gamma^{2}\sinh^2\left(\rho/\gamma\right)}-h\,.
\end{equation}
Notice that this potential is constant when $K=0$. The corresponding graph can be seen in Fig.{\ref{Fig.8}}. Observe that timelike and lightlike potentials mark off from each other just by a constant that shifts the entire curve.

This is a scattering potential where no other behaviour is possible because of the infinite asymptotic barrier caused by the diverging term. We can infer that gravitational signals or particles already in the bulk cannot reach the brane and those originated on the brane cannot travel far from it.

\subsection{BTZ with Angular Momentum}
\begin{figure}[h!]
\centering
\includegraphics[width=6.5cm]{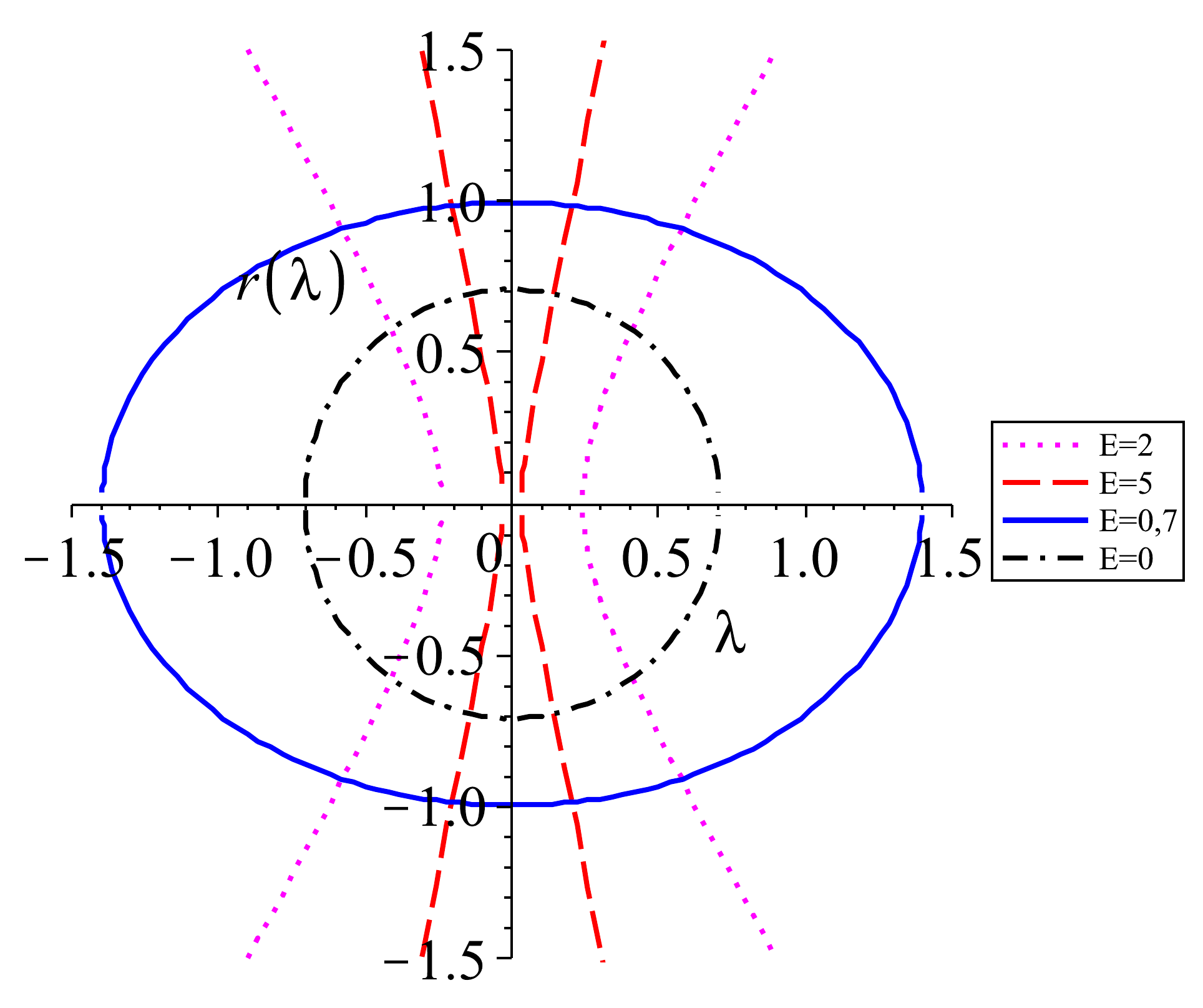}
\includegraphics[width=6.5cm]{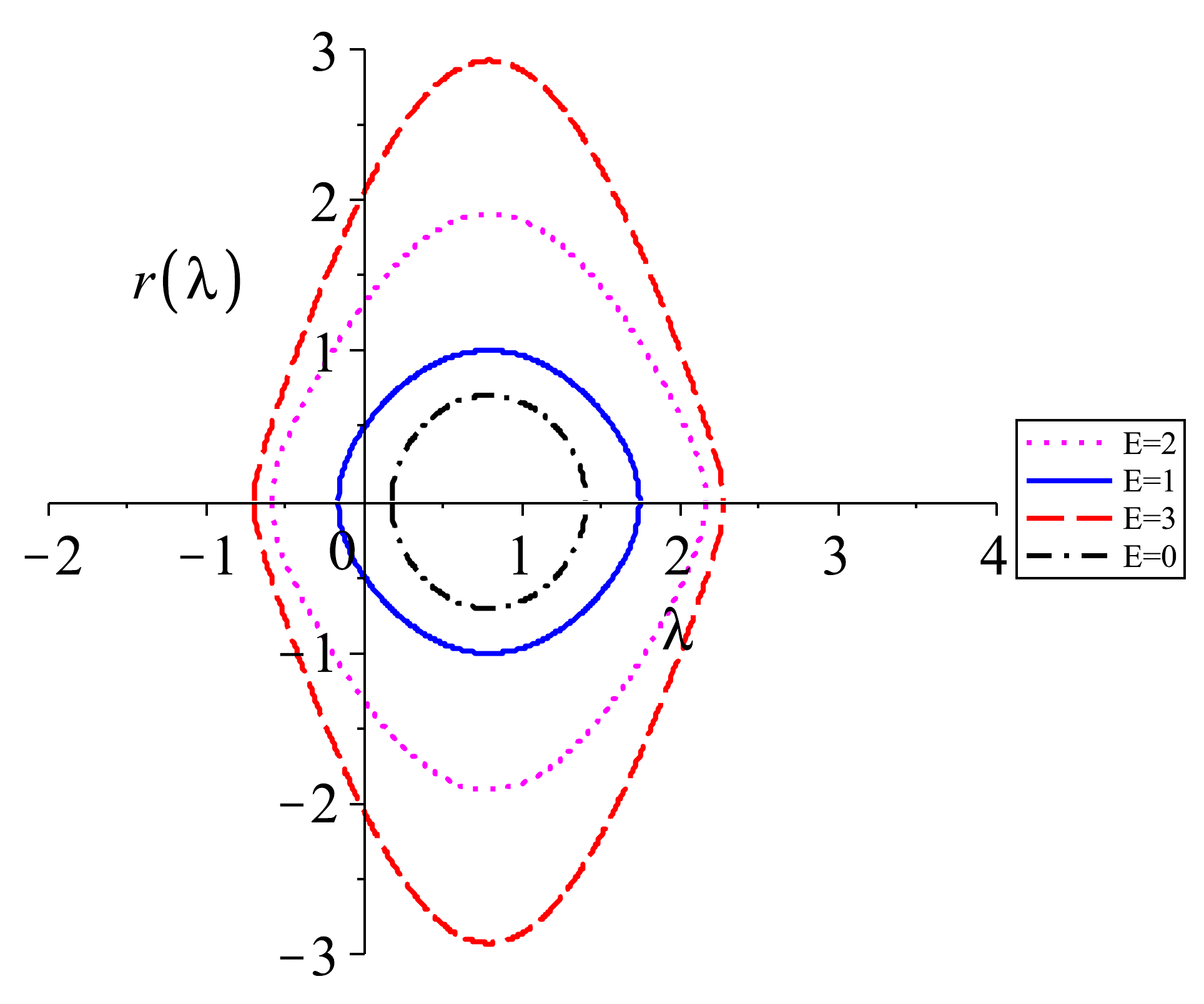}
\caption{Lightlike (left) and timelike (right) brane geodesics for BTZ solution with angular momentum. Both positive and negative branches are plotted. The energies of each geodesic are displayed on the legend. Notice that in both cases geodesics can cross the horizon, some of them after a roundabout and some others directly. Only in the lightlike case particles with high energies can escape from the black hole allowing energy extraction.}
\label{Fig.10}
\end{figure}

In this subsection we study the particular case of geodesics on the brane and in the bulk for the BTZ black hole with angular momentum in co-dimension 2-brane. The Lagrangian can be written as follows,
\begin{eqnarray}
{\cal L}&=&f^{2}\left[-\frac{1}{4}\left(-4M+\frac{4r^{2}}{l^{2}}+\frac{J^{2}}{r^{2}}\right)\dot{t}^{2}+\frac{4\dot{r}^{2}}{(-4M+4r^{2}/l^2+J^2/r^2)} \right.\nonumber \\
&&\left.+r^{2}\left(-\frac{1}{2}\frac{J\dot{t}}{r^2}+\dot{\phi}\right)^2\right]+\dot{\rho}^{2}+b^{2}\dot{\theta}^{2}\,,
\end{eqnarray}
and the equations of motion become
\begin{eqnarray}
 \frac{\partial {\cal L}}{\partial\dot{\phi}}&=&2f^{2}r^{2}(-\frac{1}{2}\frac{J\dot{t}}{r^{2}}+\dot{\phi}) = 2L \,,\\
 \frac{\partial {\cal L}}{\partial\dot{t}}&=&f^{2}\left[-\frac{1}{2}\left(-4M+\frac{4r^{2}}{l^{2}}+\frac{J^{2}}{r^2}\right)\dot{t}-\left(-\frac{1}{2}\frac{J\dot{t}}{r^2}+\dot{\phi}\right)J\right] = -2E \,,\\
 \frac{\partial {\cal L}}{\partial\dot{\theta}}&=& 2b^{2}\dot{\theta} = 2K\,.
\end{eqnarray}
With these equations the Lagrangian becomes,
\begin{equation}\label{lagf1}
h= -\frac{1}{f^2}\,\frac{E^2-LJE/r^2+ML^2/r^2-L^2/l^2}{-M+r^2/l^2+J^2/4r^2} + \frac{f^2 \dot r^2}{-M+r^2/l^2+J^2/4r^2} + \dot\rho^2 +\frac{K^2}{b^2} \,.
\end{equation}
First, we look for the geodesics on the brane. Solving this equation for $\dot{r}$ we obtain,
\begin{eqnarray}
\dot{r} &=& -\frac{1}{f^2} \left( -M  +\frac{r^2}{l^2} +\frac{J^2}{4r^2} \right) \left( \frac{K^2}{b^2} + \dot \rho^2 - h \right) \nonumber \\
&&+ \frac{1}{f^4} \left( E^2 - \frac{LJE}{r^2} + \frac{ML^2}{r^2} - \frac{L^2}{l^2} \right) \,.
\end{eqnarray}
In this case we will go through the integration to find the orbits directly. As the brane is located at $\rho=0$, $f(0)=1$ in any solution. In addition, we set $K=0$ to avoid singularities. Thus, the integral for the orbit becomes,
\begin{equation}
\lambda-\lambda_{0}=\int\frac{dr}{\sqrt{(-M+r^2/l^2+J^2/4r^2)h + (E^2-LJE/r^2+ML^2/r^2-L^2/l^2)}} \, .
\end{equation}
For the timelike case $(h=-1)$ and setting $\lambda_{0}=0$ the orbits are given by
\begin{eqnarray}
r(\lambda) &=& \pm \left[\frac{\sin(2\lambda/l)}{2} \sqrt{(Ml^2+L^2+E^2l^2)^2 - (2ELl+lJ)^2} \right. \nonumber \\
&& \left.\quad + \frac{l^2}{2}\left( M+E^2-\frac{L^2}{l^2}\right)\right]^{1/2}\,.
\end{eqnarray}
For the lightlike case ($h=0$) we have
\begin{equation}
r(\lambda) = \pm\left[\left(E^2 -\frac{L^2}{l^2}\right)\lambda^2 - \frac{(ML^2-LJE)}{E^2-L^2/l^2}\right]^{1/2}\,.
\end{equation}
The possible orbits are displayed in Fig.\ref{Fig.10}. These graphs show that geodesics can cross the event horizon in both lightlike and timelike cases. According to the energy of the particle, it can fall directly or after a roundabout in the region exterior to the horizon. Remark that the circular orbits corresponding to $E=0$ wind inside the ergosphere while the other orbits with $E>0$ extend farther. In addition, massless particles with high energies can escape from the gravitational attraction of the black hole allowing energy extraction, a behaviour already noticed in the pure ($2+1$) BTZ case~\cite{cmp}.

\bigskip

\begin{figure}[h!]
\centering
\includegraphics[width=6.3cm]{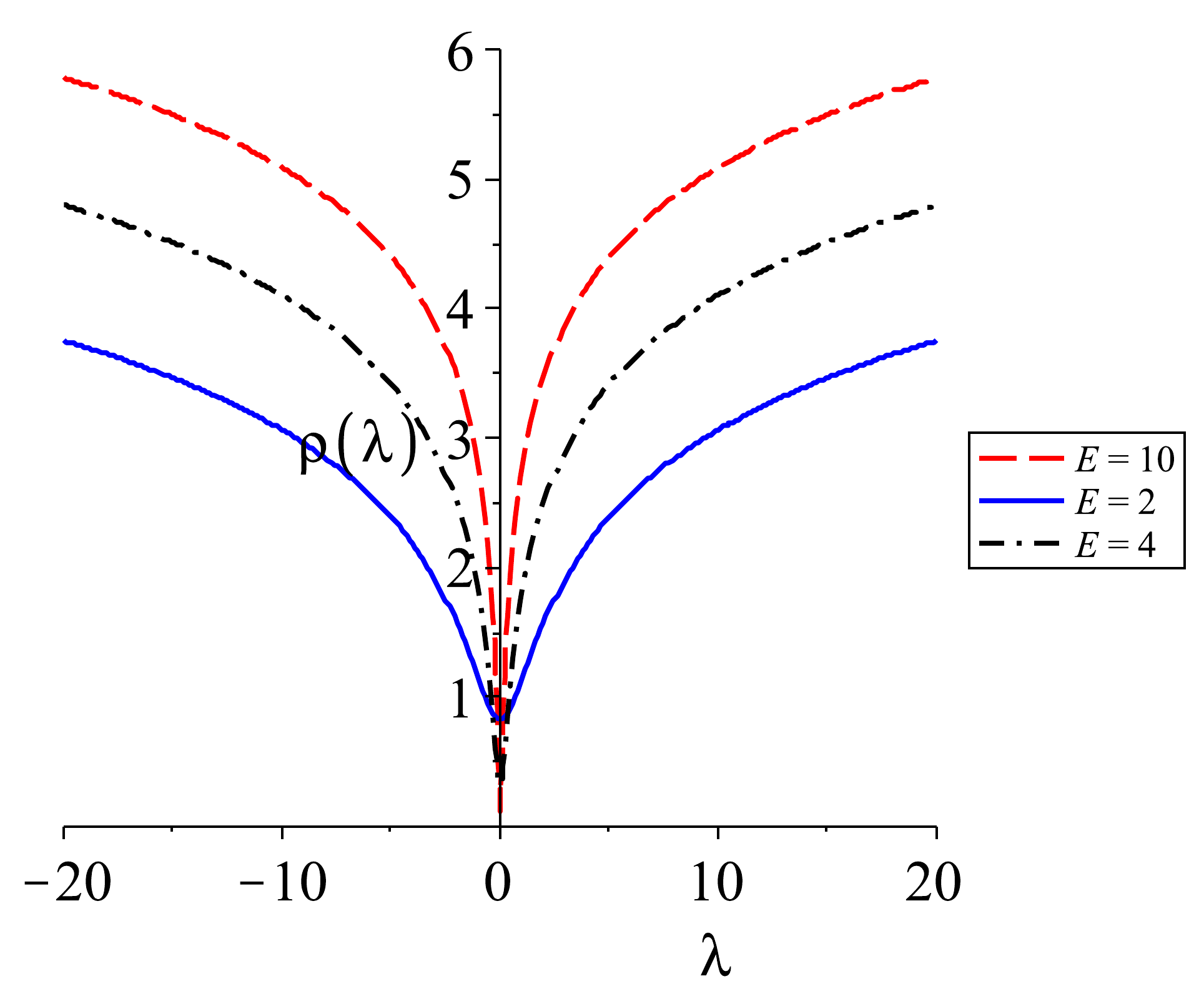}
\includegraphics[width=6cm]{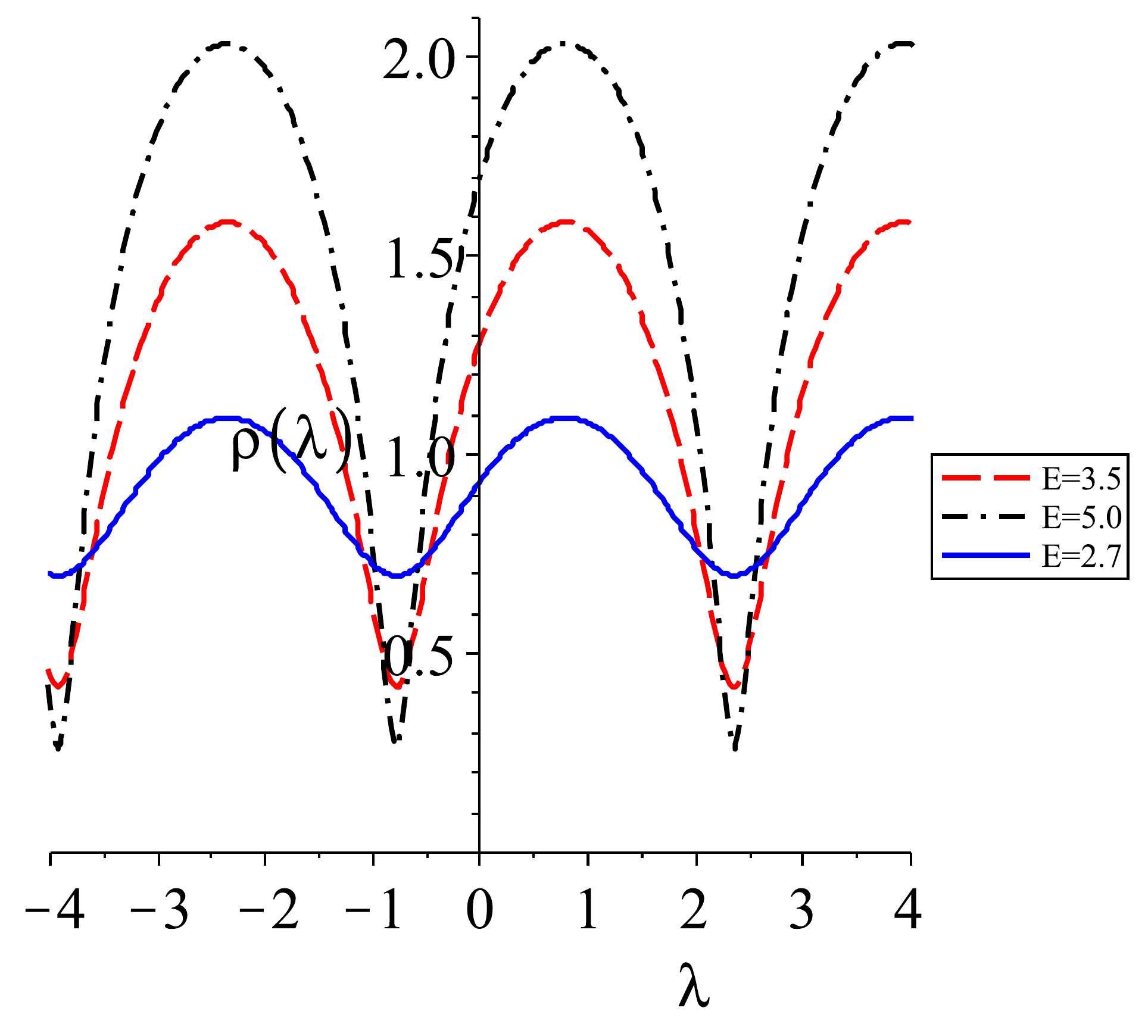}  \\
\includegraphics[width=6.3cm]{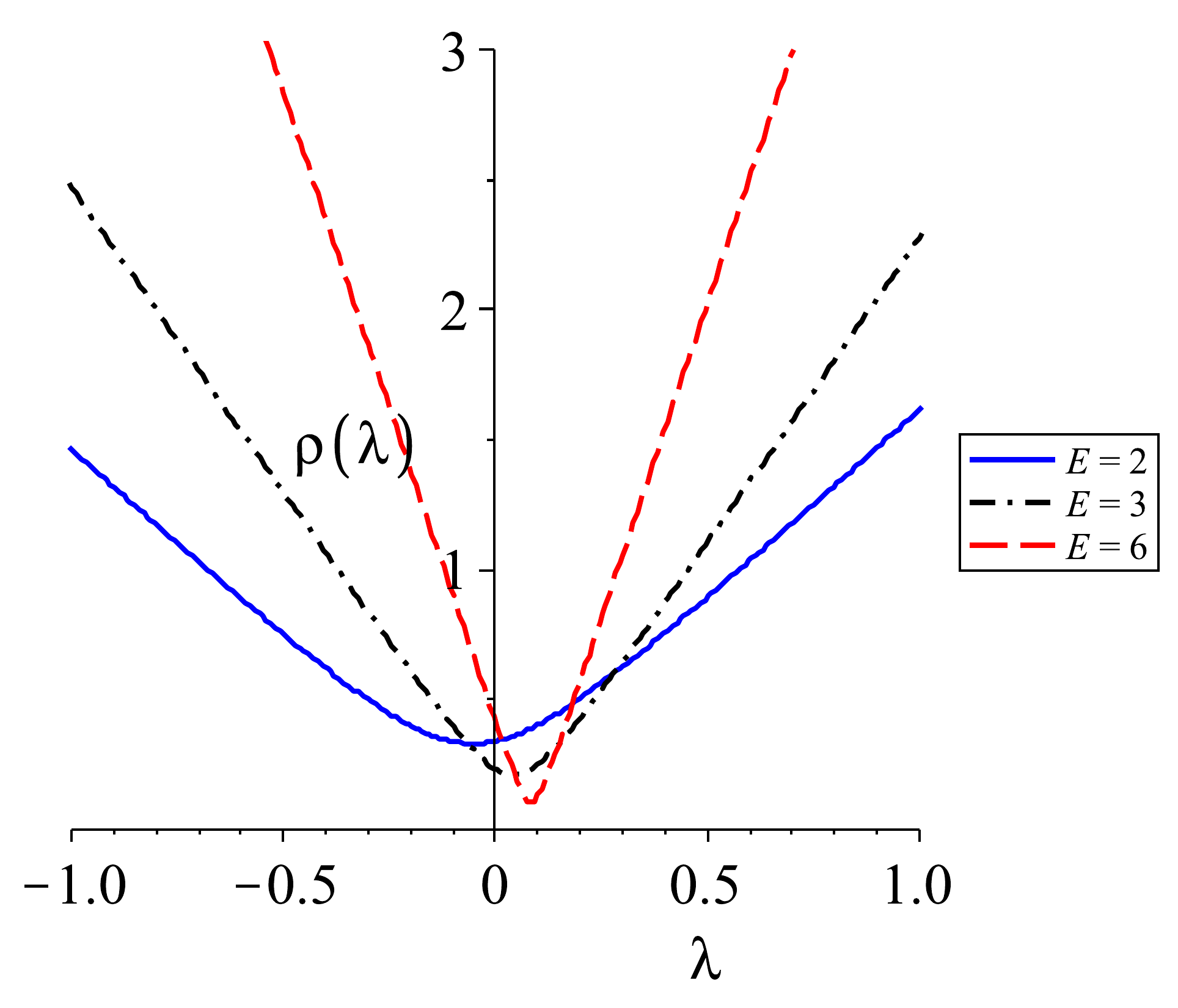}
\includegraphics[width=6cm]{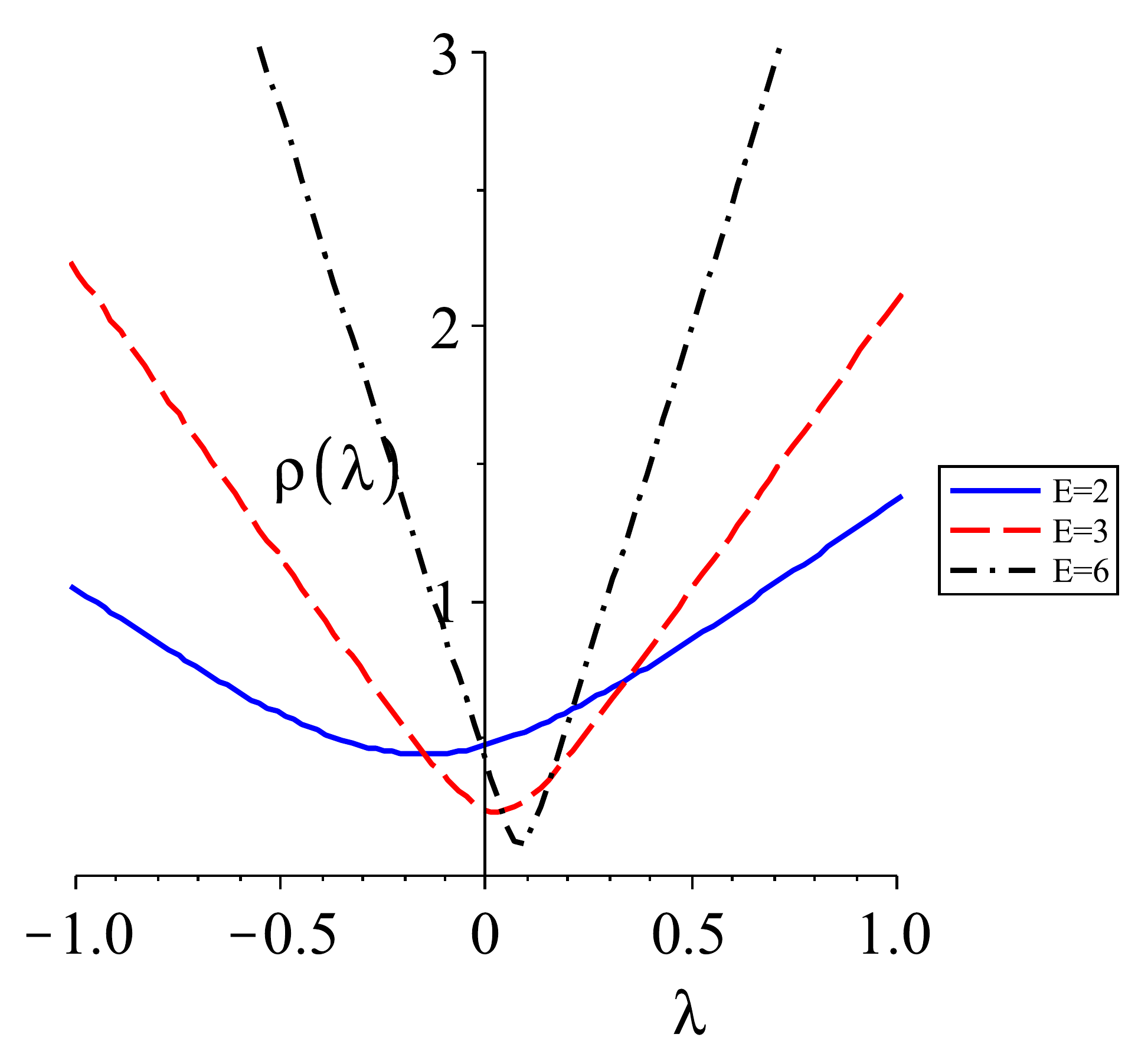}
\caption{Lightlike (left) and timelike (right) geodesics in the bulk for BTZ with angular momentum case with $f=\cosh(\rho/2\sqrt{\alpha})$ (up) and $f=1$ (down). In both cases we observe that particles cannot reach the brane. We can infer that the brane acts as a repulsive barrier both confining particles already on the brane and preventing the entrance of bulk particles.}
\label{Fig.11}
\end{figure}

Now we turn to the geodesic motion in the bulk. Accordingly, we write
Eq.(\ref{lagf1}) taking $\dot{r}=0$ and $r=\sqrt{2Ml^{2}+2l\sqrt{M^{2}l^{2}-J^{2}}} > r_E$, where $r_E$ is the ergosphere position. We first check the case $f(\rho) =\cosh(\rho/2\sqrt\alpha)$, and $b(\rho)=2\beta\sqrt{\alpha}\sinh(\rho/2\sqrt\alpha)$. After integrating, the orbits turn out to be
\begin{eqnarray}
\rho (\lambda) ^L&=& 2\sqrt{\alpha}\; \hbox{arcsinh}\left(\frac{1}{2}\sqrt\frac{4C_2\alpha+\lambda^2 C_1 ^2}{C_1 \alpha}\right) \\
\rho (\lambda) ^T&=& 2\sqrt{\alpha}\; \hbox{arcsinh}\left[\frac{1}{\sqrt{2}}\sqrt{C_1+\sqrt{C_1 ^2-4C_2}\,\sin\left(\frac{\lambda}{\sqrt\alpha}\right)}\right]\,,
\end{eqnarray}
for lightlike and timelike geodesics, respectively. The constants $C_1$ and $C_2$ are given by
\begin{equation}
C_1 = h-C_2+\frac{4}{3}\left[\frac{2l\sqrt{M^2 l^2-J^2}(E^2 -L^2/l^2)+2 l^2 E^2 M -L^2 M -J L E}{8 Ml(Ml +\sqrt{M^2 l^2-J^2})-5J^2}\right]
\end{equation}
\begin{equation}
C_2 = \frac{K^2}{4\beta^2\alpha}\,.
\end{equation}
Some trajectories are shown in Fig.\ref{Fig.11} (upper graphs). 
In both cases although the geodesics approach the brane, they cannot cross it. Analogously to the non-rotating case, we see that the brane acts like a repulsive barrier that blocks the exchange of signals between brane and bulk. Nevertheless, whereas lightlike geodesics are scattered when trying to reach the brane, timelike trajectories oscillate about a fix distance from the brane.

Similarly, the trajectories corresponding to the solutions $f(\rho)=1$, $b(\rho)=\gamma\sinh(\rho/\gamma)$ and  $f(\rho)=1$, $b(\rho)=2\beta \sqrt\alpha \sinh(\rho/2\sqrt\alpha)$ undergo an analogous pattern displayed in Fig.\ref{Fig.11} (lower graphs).
Again we see a repulsive barrier behaviour at the position of the brane that scatters lightlike and timelike particles. This prevents both signals originated on the brane to permeate the bulk and signals already in the bulk to arrive to the brane.

\section{Conclusions}

In this paper we considered the black hole solutions of five dimensional gravity with a Gauss-Bonnet term in the bulk and an induced gravity term on a 2-brane of codimension-2~\cite{CuadrosMelgar:2007jx}.
In order to explore the possibility of having more solutions to the Einstein equations we applied the Kerr-Shild method using as a background the above-mentioned solutions. This method adds to the original metric a term  depending on a null geodesic vector and a scalar function $H$. Due to the presence of the GB term the resulting equations are at most quadratic in $H$, but still solvable by choosing the appropriate geodesic vector. Working out the Einstein equations for the function $H$ we found additional solutions, which include charge, angular momentum, and brane scalar fields coupled to the brane black hole. The corresponding brane energy-momentum tensors were also computed. These results lead us to deduce that the solution with free $n(r)$ found in ~\cite{CuadrosMelgar:2007jx} does not have any constraint of diagonality.

Furthermore, we studied the geodesic behaviour in the background of the original and new solutions. For this purpose we solved the Euler-Lagrange equations for the variational problem associated with the metric. Among our results we can distinguish different cases according to the energy of the particle, the associated effective potential, and the components of the angular momentum.

In the case of geodesics on the brane we can discriminate two main cases, timelike and lightlike geodesics. 
The timelike orbits, in general, display oscillations that can become unstable making a particle cross the event horizon and never return. Nevertheless, the charged BTZ string-like solution shows an additional behaviour. When the charge is small, the paths are the same as those appearing in uncharged or scalar field coupled solutions; however, as the charge grows, the minimum of the effective potential is shifted outside the event horizon and makes possible the existence of stable oscillations or bounded orbits for particles with low or negative energy.
The lightlike geodesics exhibit particles with low energies crossing the event horizon. Nonetheless, particles with high energies can escape from the black hole and, thus, allow the extraction of energy, a result already noticed in pure (2+1) BTZ black holes geodesic structure~\cite{cmp}.

Concerning the geodesics in the bulk, we fixed $r>r_{H}$ and studied two main cases, namely, $f=\cosh(\rho/2\sqrt\alpha)$ and $f=1$. In the first case, we observed that timelike trajectories corresponding to particles without angular momentum are oscillating about the position of the brane. This implies that a particle leaving the brane can return to it opening up the possibility of shortcuts, {\it i.e.}, paths that are shorter in the bulk than on the brane~\cite{sc}. The same behaviour is verified for particles having an angular momentum component along the brane ($L\not= 0$). When $K$ (the particle's angular momentum component along the bulk) is switched on, lightlike geodesics encounter a barrier-like potential around the brane. Moreover, although timelike trajectories are still oscillating and approach the brane more and more as their energy increases, they never cross it. This result prevents the exchange of bulk and brane signals, {\it i.e.}, particles living in the bulk do not enter the brane, and particles residing on the brane cannot go far into the bulk. In the case $f=1$, both timelike and lightlike particles undergo scattering near the brane, but move in straight lines far from it because the effective potential becomes constant at this far region.

The case of a BTZ black string-like object with angular momentum was treated separately due to the non-diagonal terms appearing in the metric. We found that both lightlike and timelike geodesics on the brane can cross the horizon directly or after a roundabout in the ergosphere (particles with the lowest allowed energy) or even outer regions. However, massless particles with high energies are able to escape allowing energy extraction from the black hole. With respect to bulk geodesics, we determined that regardless the expression for $f$, the effective potential has a barrier-like structure around the brane, which both yields scattering bulk geodesics and confines particles already on the brane. In particular, timelike geodesics in the background of $f=\cosh(\rho/2\sqrt\alpha)$ oscillate around a fixed distance from the brane but never make contact with it.

It would be interesting to investigate if these features remain valid in codimension-2 $(3+1)$ brane scenarios, and if these results may help uncover other interactions between the brane and the bulk.
However, as this is out of the scope of this paper, we expect to address these questions in a future work.

\section*{Acknowledgments}

B.~C-M. and S.~A. thank the hospitality of Facultad de Ciencias F\'{i}sicas y Matem\'aticas of Universidad de Chile, where this work was accomplished.

\end{document}